\documentclass[journal]{IEEEtran}
\usepackage{cite}
\ifCLASSINFOpdf
\else
\fi
\usepackage{amsmath,amssymb,amsfonts,amsthm}
\DeclareMathOperator*{\argmin}{arg\,min}
\usepackage{mathtools}

\renewcommand{\thesection}{\arabic{section}}
\renewcommand\thesubsectiondis{\thesectiondis\arabic{subsection}.} \renewcommand\thesubsubsectiondis{\thesubsectiondis\arabic{subsubsection}.} 

\usepackage{color}
{
    \theoremstyle{plain}
    \newtheorem{assumption}{Assumption}
}

{
    \theoremstyle{plain}
    \newtheorem{theorem}{Theorem}
}

{
    \theoremstyle{plain}
    \newtheorem{lemma}{Lemma}
}

\usepackage{lipsum}
\usepackage{cuted}
\usepackage[utf8]{inputenc}
\usepackage[caption=false,font=footnotesize]{subfig}
\begin{document}
%
% paper title
% Titles are generally capitalized except for words such as a, an, and, as,
% at, but, by, for, in, nor, of, on, or, the, to and up, which are usually
% not capitalized unless they are the first or last word of the title.
% Linebreaks \\ can be used within to get better formatting as desired.
% Do not put math or special symbols in the title.
\title{Multi-Scan Implementation of the Trajectory Poisson Multi-Bernoulli Mixture Filter}
%
%
% author names and IEEE memberships
% note positions of commas and nonbreaking spaces ( ~ ) LaTeX will not break
% a structure at a ~ so this keeps an author's name from being broken across
% two lines.
% use \thanks{} to gain access to the first footnote area
% a separate \thanks must be used for each paragraph as LaTeX2e's \thanks
% was not built to handle multiple paragraphs
%
% \author{Yuxuan Xia, Karl Granstr\"{o}m, Lennart Svensson}
% \author{\IEEEauthorblockN{Yuxuan Xia\IEEEauthorrefmark{2},
% Karl Granstr\"{o}m\IEEEauthorrefmark{2},
% Lennart Svensson\IEEEauthorrefmark{2}}
% \IEEEauthorblockA{\IEEEauthorrefmark{2}Department of Electrical Engineering, Chalmers University of Technology, G\"{o}teborg, Sweden}
% }
% \author{\IEEEauthorblockN{Yuxuan Xia, Karl Granstr\"{o}m, Lennart Svensson}
% \IEEEauthorblockA{Department of Electrical Engineering, Chalmers University of Technology, G\"{o}teborg, Sweden\\
% Email: firstname.lastname@chalmers.se}}

\author{Yuxuan Xia, Karl Granstr\"{o}m, Lennart Svensson, \'{A}ngel F. Garc\'{i}a-Fern\'{a}ndez and Jason L. Williams% <-this % stops a space
\thanks{Manuscript received January 2, 2019; revised May 6, 2019 and September 3, 2019; accepted for publication November 14, 2019. \newline \indent Authors' addresses: Y. Xia, K. Granstr\"{o}m and L. Svensson, the Department of Electrical Engineering, Chalmers University of Technology, G\"{o}teborg, Sweden. \'{A}. F. Garc\'{i}a-Fern\'{a}ndez, the Department of Electrical Engineering and Electronics, University of Liverpool,
Liverpool, United Kingdom. J. L. Williams, the Commonwealth Scientific and Industrial Research Organization and Queensland University of Technology, Brisbane, Australia. \newline \indent Refereeing of this contribution was handled by Chee-Yee Chong.}}% <-this % stops a space

\maketitle

% As a general rule, do not put math, special symbols or citations
% in the abstract or keywords.
\begin{abstract}
% An analytic solution to the multi-trajectory Bayes recursion known as the Poisson multi-Bernoulli mixture (PMBM) trajectory filter has been recently proposed in \cite{continuityPMBM}. As a sequel to \cite{continuityPMBM}, this paper proposes an efficient implementation of the PMBM trajectory filter, in each iteration of which, the most likely global hypothesis is approximately obtained by solving a multi-frame assignment problem. 

The Poisson multi-Bernoulli mixture (PMBM) and the multi-Bernoulli mixture (MBM) are two multi-target distributions for which closed-form filtering recursions exist. The PMBM has a Poisson birth process, whereas the MBM has a multi-Bernoulli birth process. This paper considers a recently developed formulation of the multi-target tracking problem using a random finite set of trajectories, through which the track continuity is explicitly established. A multi-scan trajectory PMBM filter and a multi-scan trajectory MBM filter, with the ability to correct past data association decisions to improve current decisions, are presented. In addition, a multi-scan trajectory $\text{MBM}_{01}$ filter, in which the existence probabilities of all Bernoulli components are either 0 or 1, is presented. This paper proposes an efficient implementation that performs track-oriented $N$-scan pruning to limit computational complexity, and uses dual decomposition to solve the involved multi-frame assignment problem. The performance of the presented multi-target trackers, applied with an efficient fixed-lag smoothing method, are evaluated in a simulation study.
\end{abstract}

% Having observed that the hypothesis structure in the PMBM model resembles the one in the track-oriented multiple hypothesis tracker (TO-MHT), we use approximation techniques that are widely used in the TO-MHT, e.g., Lagrangian relaxation, N-scan pruning, to obtain a tractable trajectory filter with PMBM recursions.

% Note that keywords are not normally used for peerreview papers.

% For peer review papers, you can put extra information on the cover
% page as needed:
% \ifCLASSOPTIONpeerreview
% \begin{center} \bfseries EDICS Category: 3-BBND \end{center}
% \fi
%
% For peerreview papers, this IEEEtran command inserts a page break and
% creates the second title. It will be ignored for other modes.
\IEEEpeerreviewmaketitle

\section{\normalfont Introduction}
Multi-target tracking (MTT) refers to the problem of jointly estimating the number of targets and their trajectories from noisy sensor measurements \cite{mtt}. The number of targets and their trajectories can be time-varying due to targets appearing and disappearing. In a general MTT system, a multi-target tracker needs to tackle the modeling of births and deaths of targets, as well as the partitioning of noisy sensor measurements into potential tracks and false alarms; the latter is also referred to as data association. The major approaches to MTT include the joint probabilistic data association (JPDA) filter \cite{jpda}, the multiple hypothesis tracker (MHT) \cite{blackman1999design,bar2011tracking,blackman2004multiple} and random finite sets (RFS) \cite{rfs} based multi-target filters \cite[Chap. 6]{challa2011fundamentals}.

The JPDA filter \cite{jpda} seeks to calculate the marginal distribution of each track. To accommodate for an unknown and time-varying number of targets, the joint integrated probabilistic data association (JIPDA) \cite{musicki2004joint} extends the basic JPDA \cite{jpda} by incorporating target existence as an additional random variable to be estimated. It has recently been shown that the marginal data association probabilities can be efficiently approximated using message passing algorithms \cite{williams2014approximate,meyer2018message}. 

% Two popular approaches to MTT are the multiple hypothesis tracker (MHT) \cite{homht} and random finite sets (RFS) based multi-target filters \cite{rfs}. The relationship between these two approaches has been discussed in \cite{mori2016three,brekke2018relationship}.

MHT is described in a number of books, e.g., see \cite[Chap. 16]{blackman1999design}, \cite[Chap. 6, 7]{bar2011tracking}. The model was made rigorous in \cite{mori1986tracking} through random finite sequences, under the assumption that the number of targets present is constant but unknown, with an a priori distribution that is Poisson. In MHT, multiple data association hypotheses are formed to explain the source of the measurements. Each data association hypothesis assigns measurements to previously detected targets, newly detected targets, or false alarms. Data association uncertainty is captured by the data hypothesis weight, and the target state uncertainty is captured by the target state density distribution conditioned on each hypothesis. 

There are two types of MHT algorithms: the hypothesis-oriented MHT (HOMHT) \cite{homht} and the track-oriented MHT (TOMHT) \cite{morefield1977application,tomht}. In HOMHT, multiple global hypotheses are formed and evaluated between consecutive time scans; the complete algorithmic approach was first developed by Reid \cite{homht}. The TOMHT operates by maintaining a number of single target hypothesis trees, each of which contains a number of single target hypotheses explaining the measurement association history of a potential target. 

A TOMHT algorithm usually uses a deferred decision logic to consider the data associations of measurements from more than one scan, in the sense that the hypotheses are propagated into the future in anticipation that subsequent data will resolve the uncertainty \cite{blackman2004multiple}. Intuitively, measurements in more than one scan may provide more accurate data association than those in a single scan. The number of single target hypotheses can be limited by performing $N$-scan pruning \cite{blackman2004multiple}, and the involved multi-frame assignment problem is typically solved using Lagrangian relaxation based methods \cite{lagrange1,lagrange2,dualdecomposition}. Track management (target initiation and termination) is usually performed using some external procedures, see, e.g., \cite{coraluppi2012modified}.

Random finite sets (RFS) and Finite Set Statistics (FISST) were developed to provide a systematic methodology for dealing with MTT problems involving a time-varying number of targets \cite{rfs}. The relationship between RFS based approaches to MTT and MHT has been discussed in \cite{mori2016three,brekke2018relationship}. In the RFS formulation of MTT, the multi-target filtering density contains the information of the target states at the current time step. Exact closed-form solutions of RFS-based multi-target Bayes filter are given by multi-target conjugate priors. The concept of multi-target conjugate prior was defined in \cite{glmbconjugateprior} as ``If we start with the proposed conjugate initial prior, then all subsequent predicted and posterior distributions have the same form as the initial prior.'' 

% Popular approaches to solving the MTT include the joint probabilistic data association filter \cite{jpda}, the multiple hypothesis tracker (MHT) \cite{homht,tomht,blackman2004multiple} and multi-target filters based on random finite sets (RFS) \cite{rfs}. 

% In the past few decades, the track-oriented MHT \cite{tomht} is the most widely used MTT algorithm, which provides a tractable approach for implementing the MHT by maintaining a hypothesis tree for each target, and approximately finding the most likely global hypothesis by solving a multi-frame assignment problem, using, e.g., Lagrangian relaxation \cite{lagrange1,lagrange2,dualdecomposition}.

Two well-established MTT conjugate priors for the standard point target measurement model are the Poisson multi-Bernoulli mixture (PMBM) \cite{pmbmpoint} based on unlabelled RFSs, and the generalized labelled multi-Bernoulli (GLMB) \cite{glmbconjugateprior} based on labelled RFSs. The PMBM consists of a Poisson distribution representing targets which are hypothesized to exist but have not been detected, and a multi-Bernoulli mixture (MBM) representing targets that have been detected at some stage. The resulting PMBM filter \cite{pmbmpoint2} is a computationally tractable filter for the standard point target dynamic model, where the birth model is a Poisson RFS. If the birth process is a multi-Bernoulli RFS, the multi-target conjugate prior is of the form multi-Bernoulli mixture (MBM) \cite{pmbmpoint2,angel2019mbm}. A discussion regarding the differences between the use of a Poisson birth model and the use of a multi-Bernoulli birth model can be found in \cite{angel2019mbm}.

\subsection{Track Continuity in MTT}
In this subsection, we discuss how track continuity can be maintained in different MTT methodologies. Vector-type MTT methods, e.g., the JPDA filter and the MHT, describe the multitarget states and measurements by random vectors. They are able to explicitly maintain track continuity, i.e., they associate a state estimate with a previous state estimate or declare the appearance of a new target \cite{meyer2018message}. For multi-target filters based on unlabelled RFS, time-sequences of tracks cannot be constructed easily due to the set representation of the multi-target states which are order independent. The PMBM filter (as well as the MBM filter) seemingly does not provide explicit track continuity between time steps\footnote{The PMBM filter and the MBM filter are able to maintain track continuity implicitly, in a practical setting, based on information provided by meta-data.}, although a hypothesis structure in analogy to MHT was observed in \cite{pmbmpoint,pmbmpoint2}. 

One approach to addressing the lack of track continuity is to add unique labels to the target states and estimate target states from the multi-target filtering density \cite{glmbconjugateprior,garcia2013two,aoki2016labeling}. This procedure can work well in some cases but it becomes problematic in challenging situations, for example, when target birth is independent and identically distributed, and when targets get in close proximity and then separate \cite{trackingbasedontrajectories}. The $\delta$-GLMB filter \cite{glmbpoint} (and its approximation the labelled multi-Bernoulli (LMB) filter \cite{lmb}) is an example of the resulting labelled filter when the birth model is a labelled multi-Bernoulli (mixture) RFS. The $\delta$-GLMB density is similar in structure to labelled MBM using $\text{MBM}_{01}$ parameterization \cite{pmbmpoint2}, in which Bernoulli components are uniquely labelled, and their existence probability is restricted to either 0 or 1. It was shown in \cite{pmbmpoint2} that the MBM parameterization has computational and implementational advantages over the $\text{MBM}_{01}$ parameterization.

\subsection{Trajectory PMBM Filter and Its Relation to MHT}
In this subsection, we give a brief introduction to the trajectory PMBM filter and discuss its relation to MHT. More details of the trajectory PMBM filter will be given in Section \ref{para:pmbm}.

Compared to augmenting target states with unique labels, a more appealing approach to ensuring track continuity for RFSs-based multi-target filters is to generalize the concept of RFSs of targets to RFSs of trajectories. The theoretical background to perform multiple target tracking using RFS of trajectories was provided in \cite{svensson2014target,trackingbasedontrajectories}. Within the set of trajectories framework, the goal of MTT is to recursively compute the posterior density over the set of trajectories, which contains full information about the target trajectories, and can be used to estimate the best set of trajectories at each time step. 

Closed-form PMBM filtering recursions based on the sets of trajectories framework have been derived in \cite{continuityPMBM}, which enables us to leverage on the benefits of the PMBM filter recursion based on sets of targets, while also obtaining track continuity. Assuming standard point target dynamic \cite[Sec 13.2.4]{mahler2007statistical} and measurement models (defined in Section \ref{para:model}), two different trajectory PMBM filters were proposed in \cite{continuityPMBM}: one in which the set of current (i.e., alive) trajectories is tracked, and one in which the set of all trajectories (dead and alive) up to the current time step is tracked. In both cases, finite trajectories, i.e., trajectories of finite length in time, are considered. 

The implementation of the trajectory PMBM filter in \cite{continuityPMBM} considers the single-scan data association problem, and the best global hypotheses are found using Murty's algorithm \cite{murty}. As a complement to \cite{continuityPMBM}, an approximation to the exact trajectory PMBM filter that considers multi-scan data association was developed in \cite{xia2018implementation}. It operates by performing track-oriented N-scan pruning \cite{blackman2004multiple} to limit computational complexity, and using dual decomposition \cite{dualdecomposition} to solve the involved multi-frame assignment problem. The proposed algorithm therefore shares some of the key properties of certain TOMHT algorithms \cite{blackman2004multiple,dualdecomposition}, but is derived using RFSs of trajectories and birth/death models. As a comparison, TOMHT algorithms typically use heuristics to take into account the appearance and disappearance of targets \cite[Chap. 7]{bar2011tracking}.

% Similar work has been done in \cite{mori2016three}, where MHT algorithms considering multi-frame data association based on random finite set/sequence formalism are derived but with the assumption that the number of targets is unknown but constant over time. 

Numerical results in \cite{xia2018implementation} show that the proposed multi-scan trajectory PMBM filter has better tracking performance than the fast implementation of the $\delta$-GLMB filter using Gibbs sampling \cite{gibbsglmb} in terms of estimation error and computational time. These two filters use different birth models, Poisson RFS and multi-Bernoulli RFS, respectively. A multi-Bernoulli birth can be suitable if one is certain that a known maximum of targets will enter the area of interest and the targets appear around some known locations. With multi-Bernoulli birth, the PMBM conjugate prior becomes an MBM conjugate prior \cite{pmbmpoint2}. An implementation of the MBM filter for sets of targets was proposed in \cite{angel2019mbm}. The case in which the probability distribution of the number of targets is not necessarily Poisson was discussed in \cite{mori2004evaluation} for the batch-processing formulation used for TOMHT; however, a practical implementation was not provided in \cite{mori2004evaluation}.

The data association is explicitly represented in both the trajectory PMBM filter and the trajectory MBM filter, in a data structure analogous to TOMHT. Compared to conventional MHT formalism, as described in \cite{blackman2004multiple,tomht}, one important difference is that the presented trajectory PMBM filters include a Poisson RFS that models undetected trajectories. The modelling of undetected targets allows for newly discovered targets to have been born at earlier time steps \cite{brekke2018relationship}. Therefore, the trajectory PMBM filters give a higher effective birth rate than general TOMHT. The modelling of undetected targets was incorporated into TOMHT in \cite{coraluppi2014if}. In comparison, in the trajectory PMBM filters the hypotheses are purely data-to-data assignments and they are more efficiently represented using Bernoulli RFSs with probabilistic target existence. More importantly, in the PMBM trajectory filters the estimates of the set of trajectories can be directly extracted from the multi-target densities in addition to the target current states.

% two important differences are that the presented trajectory PMBM filters include a Poisson RFS that models undetected trajectories, and that the trajectory filters contain efficient representation of the data association hypotheses using Bernoulli RFSs with probabilistic target existence. The modelling of undetected targets allows for newly discovered targets to have been born at earlier time steps \cite{brekke2018relationship}. Therefore, the trajectory PMBM filter gives a higher effective birth rate than general TOMHT. The modelling of undetected targets was incorporated into TOMHT in \cite{coraluppi2014if}. In comparison, in trajectory PMBM filter the hypotheses are purely data-to-data assignments, and the hypothesis space explicitly includes the undetected targets.

\subsection{Contributions and Organization}

This paper is an extension of \cite{xia2018implementation}. In this paper, we present the trajectory PMBM and the trajectory MBM filter with multi-scan data association. The main novelties of the proposed algorithms, compared to previous work based on sets of trajectories \cite{trackingbasedontrajectories,continuityPMBM,garcia2018trajectory,garcia2018trajectorycphd}, are that they consider the multi-scan data association problem. The main novelties of the proposed algorithms, compared to TOMHT, are that they produce full trajectory estimates, i.e., smoothed estimates, upon receipt of each new set of measurements, and that the filters based on sets of trajectories model the targets that remain to be detected and the target death subsequent to the final detection. 

The contributions can be summarized as follows:
\begin{enumerate}
    \item We present the filtering recursions for the trajectory MBM filter and the trajectory $\textrm{MBM}_{01}$ filter using a multi-Bernoulli birth model. Two variants are considered for each filter: the set of current trajectories and the set of all trajectories. 
    \item We show that the ideas from the efficient TOMHT in \cite{dualdecomposition} can be utilized in trajectory filters based on PMBM, MBM and $\text{MBM}_{01}$ conjugate priors, resulting in so-called multi-scan trajectory filters. 
    \item We explain how to efficiently perform fixed-lag smoothing to extract smoothed trajectory estimates for the presented algorithms.
    \item We evaluate the performance of the presented algorithms in a simulation study, in terms of target state/trajectory estimation error and computational time.
\end{enumerate}

The paper is organized as follows. In Section II, we introduce the modeling assumption and background on sets of trajectories. In Section III, we review the PMBM conjugate prior on the set of trajectories. In Section IV, we present the filtering recursion for trajectory MBM filter. In Section V, we present implementations of the multi-scan trajectory filters. In Section VI, we present how to efficiently perform fixed-lag smoothing when extracting trajectory estimates.
Simulation results are presented in Section VII, and conclusions are drawn in Section VIII. 
% The MHT and the JPDAF are two well established MTT frameworks that have been used for decades, and studies have shown that the MHT works better than the JPDAF but with higher computational cost. There are two types of MHT, the hypothesis-oriented MHT \cite{homht} and the track-oriented MHT \cite{tomht}. The hypothesis-oriented MHT keeps a number of global hypotheses between consecutive scans whereas the track-oriented MHT only maintains a set of tracks, each of which contains a set of incompatible target hypotheses \cite{mtt}. 

\section{\normalfont Modelling}

In this section, we first outline the modeling assumptions utilized in this work. Next, we give a brief introduction to RFSs of trajectories. Then, we introduce the generalized transition and measurement models in the framework of set of trajectories; the precise mathematical definitions can be found in \cite{trackingbasedontrajectories}. The modelling is probabilistic, and the interested reader can find the necessary details about FISST, measure theory, probability generating functionals and functional derivatives for sets of trajectories in Appendices \ref{ap:a} and \ref{ap:pgfl}.

\subsection{Modeling Assumptions}
\label{para:model}

We assume that for each discrete time $k$ (a non-negative integer), a continuous time $t_k$ is assigned, such that $t_{k} > t_{k^{\prime}}$ for $k > k^{\prime}$. In the traditional formulation for RFSs of targets, target states and measurements are represented in the form of finite sets \cite{rfs}. A random single target state $x_k$ is a random element of the state (Euclidean) space $\mathcal{X} = \mathbb{R}^n$, and a random measurement $z_k$ is a random element of the measurement space $\mathcal{Z} = \mathbb{R}^m$, all at discrete time $k$. The random set of measurements obtained by a single-sensor, including clutter and target measurements with unknown origin, at time step $k$ is denoted as $\mathbf{z}_k\in\mathcal{F}(\mathcal{Z})$, where $\mathcal{F}(\mathcal{Z})$ denotes the set of all the finite subsets of $\mathcal{Z}$. 

% State and measurement spaces are both presumed to come equipped with a measure-theoretic integration concept.

We proceed by introducing two families of RFSs that will have prominent roles throughout the paper: the Poisson RFS \cite[Sec. 4.3.1]{rfs} and the Bernoulli RFS \cite[Sec. 4.3.3]{rfs}. A Poisson RFS $\boldsymbol{\Psi}$ has multi-object density distribution
\begin{equation}
    f^{\text{ppp}}(\boldsymbol{\Psi}) = e^{-\int \lambda(\Psi)d\Psi}\prod_{\Psi\in\boldsymbol{\Psi}}\lambda(\Psi),
    \label{eq:ppp}
\end{equation}
where $\lambda(\cdot)$ is the intensity function and the number of objects is Poisson distributed. An RFS $\boldsymbol{\Psi}$ is a Bernoulli RFS if $|\boldsymbol{\Psi}|\leq 1$, and a Bernoulli RFS has multi-object density distribution
\begin{equation}
    f^{\text{ber}}(\boldsymbol{\Psi}) = \begin{cases}
        1-r,& \boldsymbol{\Psi}=\emptyset\\
        rf(\Psi),& \boldsymbol{\Psi}=\{\Psi\}\\
        0,& \text{otherwise}
    \end{cases}
    \label{eq:bernoulli}
\end{equation}
where $f(\cdot)$ is a single object probability density and $r$ is the probability of existence. A multi-Bernoulli RFS is the union of a finite number of independent Bernoulli RFSs.

% In this paper, the state variable is a RFS of trajectories, as suggested in \cite{trackingbasedontrajectories}. Single trajectories are parameterized using a tuple $X = (\beta,\varepsilon,\mathbf{x}_{\beta:\varepsilon})$ \cite{continuityPMBM,svensson2014target}, where $\beta$ is the discrete time of the trajectory birth, i.e., the time the trajectory begins; $\varepsilon$ is the discrete time of the trajectory's most recent state (If $k$ is the current time, $\varepsilon<k$ means that the trajectory ends at time $\varepsilon$, and $\varepsilon=k$ means that the trajectory is ongoing); $\mathbf{x}_{\beta:\varepsilon}$ is, given $\beta$ and $\varepsilon$, the sequence of states $(x_\beta, x_{\beta+1},...,x_{\varepsilon-1},x_{\varepsilon})$.
% \begin{itemize}
%     \item $\beta$ is the discrete time of the trajectory birth, i.e., the time the trajectory begins.
%     \item $\varepsilon$ is the discrete time of the trajectory's most recent state. If $k$ is the current time, $\varepsilon<k$ means that the trajectory ends at time $\varepsilon$, and $\varepsilon=k$ means that the trajectory is ongoing.
%     \item $\mathbf{x}_{\beta:\varepsilon}$ is, given $\beta$ and $\varepsilon$, the sequence of states $(x_\beta, x_{\beta+1},...,x_{\varepsilon-1},x_{\varepsilon})$.
% \end{itemize}

In previous work \cite{trackingbasedontrajectories,continuityPMBM,garcia2018trajectory,garcia2018trajectorycphd} two different birth models have been used. In this paper we present multi-scan trajectory filter implementations for both birth models: the Poisson birth model defined in Assumption \ref{assumption:poissonbirth}; and the multi-Bernoulli birth model defined in Assumption \ref{assumption:mbbirth}. The standard point target measurement model is defined in Assumption \ref{assumption:meas}.

\begin{assumption}
The multi-target state evolves according to the following standard dynamic process with a Poisson birth model:
\begin{enumerate}
    \item New targets appear in the surveillance area independently of any existing targets. Targets arrive at each time step according to a Poisson RFS with birth intensity $\lambda^b_k(x_k)$ defined on the target state space $\mathcal{X}$. 
    % The number of targets contained in the infinitesimal region $dx_k$ centered at $x_k$ is given by $\lambda^b_k(x_k)dx_k$ on $\mathcal{X}$, and the expected number of new born targets is given by $\int \lambda^b_k(x_k)dx_k$.
    \item Given a target with state $x_k$, the target survives with a probability $P^S(x_k)$ and moves with a Markov state transition density $\pi(x_{k+1}|x_{k})$ defined on the target state space $\mathcal{X}$. The state transition density is the density of the target state at time step $k+1$, given that the target had state $x_{k}$ at time step $k$.
\end{enumerate}
\label{assumption:poissonbirth}
\end{assumption}

\begin{assumption}
The multi-target state evolves according to the following modified dynamic process with a multi-Bernoulli birth model:
\begin{enumerate}
    \item New targets appear in the surveillance area independently of any existing targets. Targets arrive at time step $k$ according to a multi-Bernoulli RFS, which has $n^b_k$ Bernoulli components. The $l$th Bernoulli component has existence probability $r^{b,l}_k$ and state density $f_k^{b,l}(x_k)$ defined on the target state space $\mathcal{X}$.
    \item Same as \textbf{Assumption \ref{assumption:poissonbirth}}, point 2.
\end{enumerate}
\label{assumption:mbbirth}
\end{assumption}

\begin{assumption}
The multi-target measurement process is as follows:
\begin{enumerate}
    \item Each target may give rise to at most one measurement, and each measurement is the result of at most one target. The probability of detection of a target with state $x_k$ is $P^D(x_k)$, and the single measurement density is $f(z_k|x_k)$ from the target space $\mathcal{X}$ to the measurement space $\mathcal{Z}$, which is the probability density of the measurement $z_k$, given that there is a target with state $x_k$ in the scene.
    \item Clutter measurements arrive according to a Poisson RFS with intensity $\lambda^{\text{FA}}(z_k)$ defined on the measurement space $\mathcal{Z}$, independently of targets and target-oriented measurements. 
    % The number of measurements contained in the infinitesimal region $dz_k$ centered at $z_k$ is given by $\lambda^{\text{FA}}(z_k)dz_k$, and the expected number of new born targets is given by $\int \lambda^{\text{FA}}(z_k)dz_k$.
\end{enumerate}
\label{assumption:meas}
\end{assumption}

% To clearly differentiate between a multi-target dynamic model with Poisson RFS birth and a multi-target dynamic model with multi-Bernoulli RFS, we refer to the former as standard point target dynamic model, and to the latter as modified point target dynamic model.

% Expressions for recursively predicting and updating the multi-trajectory filtering densities based on the above modelling assumptions can be found in \cite{trackingbasedontrajectories}. Following \cite{continuityPMBM}, we recursively express these densities as PMBMs.

    % \begin{equation}

    % \begin{align}
    %     &x_\beta, x_{\beta+1},...,x_{\varepsilon-1},x_{\varepsilon},\\
    %     & x_k \in \mathcal{X}, \forall k\in\{\beta,...,\varepsilon\}.
    % \end{align}
    % \end{equation}

% Further, let $\mathbf{X}$ denote the set of trajectories, and let $f(\mathbf{X})$ denote the corresponding density function.

\subsection{Random Finite Sets of Trajectories}

In this subsection, we first explain how the single trajectory state and its density are defined. Then, we briefly introduce some basic types of RFSs of trajectories.

% Explicit expressions for recursively predicting and updating the multi-trajectory filtering densities for a general multi-target dynamic model and the standard multi-target measurement model can be found in \cite{trackingbasedontrajectories}. Here, we first review the trajectory state representation, and then present densities for sets of trajectories.

\subsubsection{Trajectory State}

We use the trajectory state model presented in \cite{svensson2014target, trackingbasedontrajectories}, in which the trajectory state is a tuple
\begin{equation}
    X = (\beta,\varepsilon,x_{\beta:\varepsilon}),
\end{equation}
where $\beta$ is the discrete time of the trajectory birth, i.e., the time the trajectory begins; $\varepsilon$ is the discrete time of the trajectory's end time. If $k$ is the current time, $\varepsilon=k$ means that the trajectory is alive; $x_{\beta:\varepsilon}$ is, given $\beta$ and $\varepsilon$, the (finite) sequence of states
\begin{equation}
    x_{\beta:\varepsilon} =  (x_{\beta},x_{\beta+1},...,x_{\varepsilon-1},x_{\varepsilon}),
\end{equation}
where $x_{\kappa}\in\mathcal{X}$ for all $\kappa\in\{\beta,...,\varepsilon\}$. This gives a trajectory of length $l=\varepsilon - \beta + 1$ time steps. 

The single trajectory state can be considered a hybrid state consisting of discrete states $\beta$ and $\varepsilon$ representing the start and end time indices, and a continuous state $x_{\beta : \varepsilon}$ that evolves according to a stochastic model dependent on the discrete states\footnote{We remark that the use of such a hybrid state, i.e., a combination of one (or more) discrete state and one (or more) continuous state, is not uncommon in MTT: a typical example is the interacting multiple model \cite{blom1988interacting}, in which the identification of multiple models, which can be of different dimensionality \cite{granstrom2015systematic}, is governed by a discrete stochastic process.}. The trajectory state space at time step $k$ is \cite{trackingbasedontrajectories}
\begin{equation}
    \mathcal{T}_k = \uplus_{(\beta,\varepsilon)\in I_k}\{\beta\}\times\{\varepsilon\}\times\mathcal{X}^{\varepsilon-\beta+1},
    \label{eq:trajectoryspace}
\end{equation}
where $\uplus$ denotes the union of (possibly empty) sets that are mutually disjoint, $I_k=\{(\beta,\varepsilon):0\leq\beta\leq\varepsilon\leq k\}$ is the set of all possible start and end times of trajectories up to time step $k$, the $\mathcal{X}^l$ denotes $l$ Cartesian products of $\mathcal{X}$, i.e., the Cartesian products of spaces of different sizes. A trajectory state density $p(\cdot)$ of $X$ factorizes as follows
\begin{equation}
    p(X) = p(x_{\beta:\varepsilon}|\beta,\varepsilon)P(\beta,\varepsilon),
    \label{eq:tradensity}
\end{equation}
where, if $\varepsilon < \beta$, then $P(\beta,\varepsilon)$ is zero. Integration for single trajectory densities is performed as follows \cite{trackingbasedontrajectories},
\begin{multline}
    \int p(X)dX =\\\sum_{(\beta,\varepsilon)\in I_k}\left[\int...\int p(x_{\beta:\varepsilon}|\beta,\varepsilon)dx_{\beta}...dx_{\varepsilon}\right]P(\beta,\varepsilon).
\end{multline}

\subsubsection{Sets of Trajectories}

A set of trajectories is denoted as $\mathbf{X}_k\in\mathcal{F}(\mathcal{T}_k)$, where $\mathcal{F}(\mathcal{T}_k)$ is the set of all the finite subsets of $\mathcal{T}_k$. Let $g(\mathbf{X}_k)$ be a real-valued function on a set of trajectories, then the set integral is
\begin{multline}
    \int g(\mathbf{X}_k)\delta\mathbf{X}_k \triangleq \\ g(\emptyset) + \sum_{n=1}^{\infty}\frac{1}{n!}\int...\int g(\{X^1_k,...,X^n_k\})dX^1_k...dX_k^n.
    \label{eq:setintegral}
\end{multline}
% We explain how to define measure theoretic integrals for single object LCHS spaces in Appendix B, and we show how to choose a reference measure for sets of trajectories in Appendix C.

% Two basic building blocks for the conjugate priors are the Poisson RFS and the Bernoulli RFS. 
A trajectory Poisson RFS has (multi-trajectory) density of the form (\ref{eq:ppp}),
% \begin{equation}
%     f^{\text{ppp}}(\mathbf{X}) = e^{-\int \lambda(X)dX}\prod_{X\in\mathbf{X}}\lambda(X),
%     \label{eq:ppp}
% \end{equation}
where the trajectory Poisson RFS intensity $\lambda(\cdot)$ is defined on the trajectory state space $\mathcal{T}_k$, i.e., realizations of the Poisson RFS are trajectories with a birth time, a time of the most recent state, and a state sequence \cite{garcia2018trajectory}. A trajectory Bernoulli RFS has density of the form (\ref{eq:bernoulli}),
% \begin{equation}
%     f^{\text{ber}}(\mathbf{X}) = \begin{cases}
%         1-r,& \mathbf{X}=\emptyset\\
%         rf(X),& \mathbf{X}=\{X\}\\
%         0,& \text{otherwise}
%     \end{cases}
%     \label{eq:bernoulli}
% \end{equation}
where $f(\cdot)$ is a single trajectory density (\ref{eq:tradensity}). Trajectory multi-Bernoulli RFS and trajectory MBM RFS are both defined analogously to target multi-Bernoulli RFS and target MBM RFS \cite{trackingbasedontrajectories}: a trajectory multi-Bernoulli is the disjoint union of a multiple trajectory Bernoulli RFS; trajectory MBM RFS is an RFS whose density is a mixture of trajectory multi-Bernoulli densities.

\subsection{Transition Models for Sets of Trajectories}

In the standard multi-target dynamic model with Poisson birth, see Assumption \ref{assumption:poissonbirth}, target birth at time step $k$ is modeled by a Poisson RFS, with intensity
\begin{subequations}
    \begin{align}
        \lambda^B_k(X)&=\lambda^{B,x}_k(x_{\beta:\varepsilon}|\beta,\varepsilon)\Delta_k(\varepsilon)\Delta_k(\beta),\\
        \lambda^{B,x}_k(x_{k:k}|k,k) &= \lambda^b_k(x_k),
    \end{align}
\end{subequations}
where $\Delta(\cdot)$ denotes the Kronecker delta function. In the modified multi-target dynamic model with multi-Bernoulli birth, see Assumption \ref{assumption:mbbirth}, target birth at time step $k$ is modeled by a multi-Bernoulli RFS, with the trajectory state density in the $l$th Bernoulli component
\begin{subequations}
    \begin{align}
        f^{B,l}_k(X)&=f^{B,l,x}_k(x_{\beta:\varepsilon}|\beta,\varepsilon)\Delta_k(\varepsilon)\Delta_k(\beta),\\
        f^{B,l,x}_k(x_{k:k}|k,k) &= f^{b,l}_k(x_k),
    \end{align}
    \label{eq:mbirth}
\end{subequations}
\\*
and the existence probability $r^{b,l}_k$.

We focus on two different MTT problem formulations: the set of current trajectories, where the objective is to estimate the trajectories of targets that are still present in the surveillance area at the current time; and the set of all trajectories, where the objective is to estimate the trajectories of both the targets that are still present in the surveillance area at the current time, and the targets that once were in (but have since left) the surveillance area at some previous time. The probability of survival as a function on trajectories at time step $k$ is defined as
\begin{equation}
    P^S_k(X) = P^S(x_{\varepsilon})\Delta_k(\varepsilon).
\end{equation}
The transition density for the trajectories depends on the problem formulation. 

\subsubsection{Transition Model for the Set of Current Trajectories}
The Bernoulli RFS transition density for a single potential target without birth is \begin{subequations}
    \begin{align}
        \begin{split}
        {}&f^c_{k|k-1}(\mathbf{X}|\mathbf{X}^{\prime}) =\\ &\begin{cases}
            1,& \mathbf{X}^{\prime}=\emptyset,\mathbf{X}=\emptyset\\
            1-P^S_{k-1}(X^{\prime}),& \mathbf{X}^{\prime}=\{X^{\prime}\},\mathbf{X}=\emptyset\\
            P^S_{k-1}(X^{\prime})\pi^c(X|X^{\prime}),& \mathbf{X}^{\prime}=\{X^{\prime}\},\mathbf{X}=\{X\}\\
            0,& \text{otherwise}
        \end{cases}
        \end{split}\\
        &\pi^c(X|X^{\prime}) = \pi^{c,x}(x_{\beta:\varepsilon}|\beta,\varepsilon,X^{\prime})\Delta_{\varepsilon^{\prime}+1}(\varepsilon)\Delta_{\beta^{\prime}}(\beta),\\
        &\pi^{c,x}(x_{\beta:\varepsilon}|\beta,\varepsilon,X^{\prime}) = \pi^x(x_{\varepsilon}|x^{\prime}_{{\varepsilon^{\prime}}})\delta_{x^\prime_{\beta^{\prime}:\varepsilon^{\prime}}}(x_{\beta:\varepsilon-1}),
    \end{align}
    \label{eq:transition_current}
\end{subequations}
\\*
where $\delta(\cdot)$ denotes Dirac delta function, and $X^{\prime}$ denotes the single trajectory state at time step $k-1$. In this model, $P^S(\cdot)$ is used as follows. If the target disappears, or ``dies'', then the entire trajectory will no longer be a member of the set of current trajectories. If the trajectory survives, then the trajectory is extended by one time step.
\subsubsection{Transition Model for the Set of All Trajectories}
The Bernoulli RFS transition density for a single potential target without birth is
\begin{subequations}
    \begin{align}
        \begin{split}
        {}&f^a_{k|k-1}(\mathbf{X}|\mathbf{X}^{\prime}) =\\ &\begin{cases}
            1,& \mathbf{X}^{\prime}=\emptyset,\mathbf{X}=\emptyset\\
            \pi^a(X|X^{\prime}),& \mathbf{X}^{\prime}=\{X^{\prime}\},\mathbf{X}=\{X\}\\
            0,& \text{otherwise}
        \end{cases}
        \end{split}\\
        &\pi^a(X|X^{\prime}) = \pi^{a,x}(x_{\beta:\varepsilon}|\beta,\varepsilon,X^{\prime})\pi^{\varepsilon}(\varepsilon|\beta,X^{\prime})\Delta_{\beta^{\prime}}(\beta),\\
        {}&\pi^{\varepsilon}(\varepsilon|\beta,X^{\prime}) = \begin{cases}
            1,&\varepsilon = \varepsilon^{\prime}<k-1\\
            1-P^S_{k-1}(X^{\prime}), &\varepsilon = \varepsilon^{\prime}=k-1\\
            P^S_{k-1}(X^{\prime}),& \varepsilon = \varepsilon^{\prime}+1=k\\
            0,& \text{otherwise}
            \end{cases}\\
        \begin{split}
        {}&\pi^{a,x}(x_{\beta:\varepsilon}|\beta,\varepsilon,X^{\prime}) = \\ &\begin{cases}
            \delta_{x^{\prime}_{\beta^{\prime}:\varepsilon^{\prime}}}(x_{\beta:\varepsilon}),&\varepsilon=\varepsilon^{\prime}\\
            \pi^x(x_{\varepsilon}|x^{\prime}_{\varepsilon^{\prime}})\delta_{x^{\prime}_{\beta^{\prime}:\varepsilon^{\prime}}}(x_{\beta:\varepsilon-1}).&\varepsilon=\varepsilon^{\prime}+1
        \end{cases}
        \end{split}
    \end{align}
    \label{eq:transition_all}
\end{subequations}
\\*
In this model, the interpretation of the probability of survival is that it governs whether the trajectory ends or it is extended by one more time step. However, importantly, regardless of whether or not the trajectory ends, the trajectory remains in the set of all trajectories with probability one.

The complete transition model for sets of trajectories is analogous to the complete transition model for sets of targets, by using sets of trajectories and the corresponding Bernoulli transition density for each problem formulation. Given the set $\mathbf{X}_{k-1}=\{X_{k-1}^1,...,X_{k-1}^n\}$ of trajectories at time step $k-1$, and the set $\mathbf{X}_k$ of trajectories at time step $k$ is $\mathbf{X}_k = \mathbf{X}_k^{\text{b}}\uplus\mathbf{X}_k^1\uplus...\uplus\mathbf{X}_k^n$, where $\mathbf{X}_k^{\text{b}}$, $\mathbf{X}_k^1$,..., $\mathbf{X}_k^n$ are independent sets, $\mathbf{X}_k^{\text{b}}$ is the set of newborn trajectories and $\mathbf{X}_k^i$ is the set of trajectories resulted from $X^i_{k-1}$. Using the convolution formula for multi-object densities \cite[Eq. (4.17)]{rfs}, the resulting multi-trajectory density $f(\cdot|\cdot)$ of $\mathbf{X}_k$ given $\mathbf{X}_{k-1}$ can be written as
\begin{multline}
    f(\mathbf{X}_k|\mathbf{X}_{k-1})=\sum_{\mathbf{X}_k^{\text{b}}\uplus\mathbf{X}_k^1\uplus...\uplus\mathbf{X}_k^n = \mathbf{X}_k}f_k^{\text{birth}}(\mathbf{X}_k^{\text{b}})\\\times\prod^n_{i=1}f^{\text{persist}}_{k|k-1}(\mathbf{X}_k^i|\{X_{k-1}^i\}).
\end{multline}
% and it can be written as
% \begin{multline}
%     f(\mathbf{X}_k|\mathbf{X}_{k-1}) = \sum_{\substack{\boldsymbol{\Xi}\in \mathcal{F}(\mathcal{T}_{k-1}):\\
%     \bigcup_{\boldsymbol{\Xi}^{\prime}\subseteq\mathbf{X}_{k-1}} \boldsymbol{\Xi}\subseteq \mathbf{X}_k}} f^{\text{birth}}_k\Bigg( \mathbf{X}_k \setminus \bigcup_{\boldsymbol{\Xi}^{\prime}\subseteq\mathbf{X}_{k-1}}\boldsymbol{\Xi} \Bigg)\\\times \prod_{\boldsymbol{\Xi}^{\prime}\subseteq\mathbf{X}_{k-1}}f^{\text{persist}}_{k|k-1}(\boldsymbol{\Xi}|\boldsymbol{\Xi}^{\prime}),
% \end{multline}
% \begin{multline}
%     f(\mathbf{X}_k|\mathbf{X}_{k-1}) = \sum_{\substack{\boldsymbol{\Xi}_X\in \mathcal{F}(\mathcal{T}_{k-1}):\\
%     \bigcup_{X\in\mathbf{X}_{k-1}} \boldsymbol{\Xi}_X\subseteq \mathbf{X}_k}} f^{\text{birth}}_k\Bigg( \mathbf{X}_k \setminus \bigcup_{X\in\mathbf{X}_{k-1}}\boldsymbol{\Xi}_X \Bigg)\\\times \prod_{X\in\mathbf{X}_{k-1}}f^{\text{persist}}_{k|k-1}(\boldsymbol{\Xi}_X|\{X\}),
% \end{multline}
where $f_k^{\text{birth}}(\cdot)$ is either a trajectory Poisson RFS or a trajectory multi-Bernoulli RFS, and $f_{k|k-1}^{\text{persist}}(\cdot|\cdot)$ is a Bernoulli transition
density for a single potential target without birth, with the
form $f_{k|k-1}^a(\cdot|\cdot)$ or $f_{k|k-1}^c(\cdot|\cdot)$.

% and $\boldsymbol{\Xi}_{X}$ denotes the set of surviving trajectory from trajectory $X$. If $f^{\text{persist}}_{k|k-1}$ has the form $f^c(\cdot|\cdot)$, set $\boldsymbol{\Xi}_{X}$ contains at most one element; if $f^{\text{persist}}_{k|k-1}$ has the form $f^a(\cdot|\cdot)$, set $\boldsymbol{\Xi}_{X}$ contains one and only one element.

\subsection{Single Trajectory Measurement Model}
According to the point target measurement model in Assumption \ref{assumption:meas}, the multi-object density of a target-generated measurement at time step $k$ given a set of trajectories with 0 or 1 element is Bernoulli, with the form
\begin{subequations}
    \begin{align}
        \begin{split}
        {}&\varphi_k(\mathbf{w}_k|\mathbf{X}) =\\ &\begin{cases}
            1,& \mathbf{X}=\emptyset,\mathbf{w}_k=\emptyset\\
            1-P^D_k(X),& \mathbf{X}=\{X\},\mathbf{w}_k=\emptyset\\
            P^D_k(X)\varphi(z_k|X),& \mathbf{X}=\{X\},\mathbf{w}_k=\{z_k\}\\
            0,& \text{otherwise}
        \end{cases}
        \end{split}\\
        &P^D_k(X)=P^D(x_{\varepsilon})\Delta_k(\varepsilon),\\
        &\varphi(z|X)=f(z|x_{\varepsilon}).
    \end{align}
    \label{eq:measurementmodel}
\end{subequations}
\\*
Note that trajectories that do not exist at the current time cannot be detected. The complete measurement model for sets of trajectories is similar to the measurement model for sets of targets by using the proper probability of detection and single measurement density for trajectories \cite{trackingbasedontrajectories}. Given the set $\mathbf{X}_k = \{X_k^1,...,X_k^n\}$ of trajectories at time step $k$, the set $\mathbf{z}_k$ of measurements at time step $k$ is $\mathbf{z}_k = \mathbf{w}^c_k\uplus\mathbf{w}_k^1\uplus...\uplus\mathbf{w}_k^n$, where $\mathbf{w}_k^c$, $\mathbf{w}_k^1$,..., $\mathbf{w}_k^n$ are independent sets, $\mathbf{w}^c_k$ is the set of clutter measurements and $\mathbf{w}_k^i$ is the set of measurements produced by trajectory $i$. The resulting measurement set density $f(\cdot|\cdot)$ of $\mathbf{z}_k$ given $\mathbf{X}_k$ can be written as
\begin{equation}
    f(\mathbf{z}_k|\mathbf{X}_k) = \sum_{\mathbf{w}_k^c\uplus\mathbf{w}_k^1\uplus...\uplus\mathbf{w}_k^n=\mathbf{z}_k}f^{\text{ppp}}_k(\mathbf{w}_k^c)\prod^n_i\varphi_k(\mathbf{w}_k^i|\{{X}_k^i\}).
\end{equation}

% and it can be written as
% \begin{multline}
%     f(\mathbf{z}_k|\mathbf{X}_k) = \sum_{\substack{\substack{\mathbf{w}_k\in\mathcal{F}(\mathcal{Z}):\\\bigcup_{\boldsymbol{\Xi}\subseteq\mathbf{X}_k}\mathbf{w}_k\subseteq \mathbf{z}_k}}} f^{\text{FA}}_k(\mathbf{z}_k\setminus \bigcup_{\boldsymbol{\Xi}\subseteq\mathbf{X}_k}\mathbf{w}_k)\\\times \prod_{\boldsymbol{\Xi}\subseteq\mathbf{X}_k}\varphi_k(\mathbf{w}_k|\boldsymbol{\Xi}),
% \end{multline}
% \begin{multline}
%     f(\mathbf{W}_k|\mathbf{X}_k) = \sum_{\substack{\substack{\mathbf{Z}_X\in\mathcal{F}(\mathcal{Z}):\\\bigcup_{X\in\mathbf{X}_k}\mathbf{Z}_X\subseteq \mathbf{W}_k}}} f^{\text{clutter}}_k(\mathbf{W}_k\setminus \bigcup_{X\in\mathbf{X}_k}\mathbf{Z}_k)\\\times \prod_{X\in\mathbf{X}_k}\varphi(\mathbf{Z}_k|\{X\}),
% \end{multline}
% where $f^{\text{FA}}(\cdot)$ is the clutter Poisson RFS density.

% $\mathbf{Z}_X$ denotes the set of measurements (contains at most one element) associated to trajectory $X$ at time $k$, and 

\section{\normalfont Trajectory PMBM Filter}
\label{para:pmbm}

The PMBM conjugate prior was developed for point targets in \cite{pmbmpoint} and for extended targets in \cite{granstrom2019poisson}, and it was further generalized to trajectories in \cite{continuityPMBM,yuxuan2019pmbm}. Given the sequence of measurements up to time step $k^{\prime}$ and Assumptions 1 and 3, the density of the set of trajectories at time step $k\in\{k^{\prime},k^{\prime}+1\}$ is given by the PMBM density of the form
\begin{subequations}
    \begin{align}
        f_{k|k^{\prime}}(\mathbf{X}_k) &= \sum_{\mathbf{X}_k^u\uplus\mathbf{X}_k^d=\mathbf{X}}f_{k|k^{\prime}}^{\textrm{ppp}}(\mathbf{X}_k^u)\sum_{a\in\mathcal{A}_{k|k^{\prime}}}w^a_{k|k^{\prime}}f_{k|k^{\prime}}^a(\mathbf{X}_k^d),\\
        f_{k|k^{\prime}}^{\textrm{ppp}}(\mathbf{X}^u_k) &= e^{-\int \lambda_{k|k^{\prime}}^u(X)dX}\prod_{X\in\mathbf{X}_k^u}\lambda_{k|k^{\prime}}^u(X),\\
        f_{k|k^{\prime}}^a(\mathbf{X}_k^d) &= \sum_{\uplus_{i\in\mathbb{T}_{k|k^{\prime}}}\mathbf{X}_k^i=\mathbf{X}_k^d}\prod_{i\in\mathbb{T}_{k|k^{\prime}}}f_{k|k^{\prime}}^{i,a^i}(\mathbf{X}^i_k),
        \label{eq:mbmdensity}
    \end{align}
    \label{eq:pmbm}%
\end{subequations}
where the RFS of trajectories $\mathbf{X}_k$ is an independent union of a Poisson RFS $\mathbf{X}_u^k$ with intensity $\lambda^u_{k|k^{\prime}}$ and an MBM RFS $\mathbf{X}_k^d$ with Bernoulli parameters $r^{i,a^i}_{k|k^{\prime}}$ and $f^{i,a^i}_{k|k^{\prime}}(\cdot)$, cf. (\ref{eq:bernoulli}), and $\mathcal{A}_{k|k^{\prime}}$ is the set of all global hypotheses, which will be explained in the next subsection. A trajectory PMBM RFS can be defined by the parameters of the density,
\begin{subequations}
    \begin{align}
        &\lambda^u_{k|k^{\prime}}, \mathcal{A}_{k|k^{\prime}},\left\{ \Theta_{k|k'}^{a} \right\}_{ a \in \mathcal{A}_{k|k'}},\\
        &\Theta_{k|k'}^{a} = \left\{ \left( w_{k|k'}^{i,a^i} , r_{k|k'}^{i,a^i} , f_{k|k'}^{i,a^i} \right) \right\}_{i \in \mathbb{T}}.
    \end{align}
    \label{eq:pmbm_parameters}
\end{subequations}
% \begin{equation}
%     \lambda^u_{k|k^{\prime}}, \mathcal{A}_{k|k^{\prime}}, \{(w^{i,a^i}_{k|k^{\prime}},r^{i,a^i}_{k|k^{\prime}},f^{i,a^i}_{k|k^{\prime}})\}.
%     \label{eq:pmbm_parameters}
% \end{equation}

% The PMBM density is defined entirely by the parameters of the density
% \begin{equation}
% \lambda^u, \{(\mathcal{W}^j,\{r^{j,i},f^{j,i}(\cdot)\}_{i\in\mathbb{I}^j})\}_{j\in\mathbb{J}}.
% \label{eq:update}
% \end{equation}
% Because the PMBM density is a MTT conjugate prior, performing prediction (\ref{eq:bayespredict}) and update (\ref{eq:bayesupdate}) means that we compute the predicted and updated, respectively, PMBM density parameters.

\subsection{Structure of the Trajectory PMBM Filter}
The structure of the trajectory PMBM (\ref{eq:pmbm}) is in analogy to the structure of the target PMBM \cite{pmbmpoint}. The Poisson RFS represents trajectories that are hypothesized to exist, but have never been
detected, i.e., no measurement has been associated to them. In the track-oriented trajectory PMBM filter, a new track is initiated for each measurement received. In the MBM in (\ref{eq:pmbm}), $\mathbb{T}_{k|k^{\prime}}=\{1,...,n_{k|k^{\prime}}\}$ is a track table with $n_{k|k^{\prime}}$ tracks, $a=(a^1,...,a^{n_{k|k^{\prime}}})\in\mathcal{A}_{k|k^{\prime}}$ is a possible global data association hypothesis, and for each global hypothesis $a$ and for each track
$i\in\mathbb{T}_{k|k^{\prime}}$, $a^i$ indicates which track hypothesis is used in the global hypothesis. For each track, there are $h^i_{k|k^{\prime}}$ single trajectory hypotheses\footnote{The ``track'' defined here is different from the convention used in MHT algorithms, where ``track'' is referred to as single trajectory hypothesis.}. The weight of global hypothesis $a$ is $w^a_{k|k^{\prime}}\propto\prod_{i\in\mathbb{T}_{k|k^{\prime}}}w^{i,a^i}_{k|k^{\prime}}$, where $w^{i,a^i}_{k|k^{\prime}}$ is the weight of single trajectory hypothesis $a^i$ from track $i$.

Let $m_k$ be the number of measurements at time step $k\in\{1,...,\tau\}$ and $j\in\mathbb{M}_k=\{1,...,m_k\}$ be an index to each measurement. Let $\mathcal{M}_{k}$ denote the set of all measurement indices up to and including time step $k$; the elements of $\mathcal{M}_{k}$, if not empty, are of the form $(\tau,j)$, where $j\in\{1,...,m_{\tau}\}$ is an index of a measurement at time step $\tau\leq k$. Further, let $\mathcal{M}^k(i,a^i)$ denote the history of measurements that are hypothesized to belong to hypothesis $a^i$ from track $i$ at time step $k$. Under the standard point target measurement model assumption, see \textbf{Assumption \ref{assumption:meas}}, there can be at maximum one measurement corresponding to the same time step in $\mathcal{M}^k(i,a^i)$.

For a global hypothesis to be correct, we have the following constraints. Each global hypothesis should explain the association of each measurement received so far. In addition, every measurement should be associated to one and only one track in each global hypothesis. In other words, the single trajectory hypotheses included in a given global hypothesis cannot have any shared measurement. Under these constraints, the set of global hypotheses at time step $k$ can be expressed as
\begin{multline}
    \mathcal{A}_{k|k^{\prime}}=\bigg\{a=(a^1,...,a^{n_{k|k^{\prime}}})\bigg|\bigcup_{i\in\mathbb{T}_{k|k^{\prime}}}\mathcal{M}^k(i,a^i)=\mathcal{M}_k,\\\mathcal{M}^k(i,a^i)\cap\mathcal{M}^k(j,a^j)=\emptyset~\forall~i\neq j,~i,j\in\mathbb{T}_{k|k^{\prime}}\bigg\}.
    \label{eq:globalhypothesis}
\end{multline}

% \begin{multline}
%     \mathcal{A}_{k|k^{\prime}}=\bigg\{(\mathbf{a}^1_{k|k^{\prime}},...,\mathbf{a}^{n_{k|k^{\prime}}}_{k|k^{\prime}})\bigg|\mathbf{a}^i_{k|k^{\prime}}\in\mathcal{A}^i_{k|k^{\prime}}, \bigcup_{i\in\mathbb{X}_{k|k^{\prime}}}\mathcal{M}(\mathbf{a}^i_{k|k^{\prime}})\\=\mathcal{M}_{k^{\prime}}, \mathcal{M}(\mathbf{a}^i_{k|k^{\prime}})\cap\mathcal{M}(\mathbf{a}^j_{k|k^{\prime}})=\emptyset~\forall~i\neq j\bigg\}.
%     % \label{eq:globalhypothesis}
% \end{multline}

\subsection{PMBM Filtering Recursion}

The form of the PMBM conjugate prior on the sets of trajectories is preserved through prediction and update. The two different trajectory PMBM filters based on the two different transition models for sets of trajectories are both track-oriented. For each track, there is a hypothesis tree, where each hypothesis corresponds to different data association sequences for the track. The prediction step preserves the number of tracks and the number of hypotheses. By using a Poisson RFS birth model, the density of new born trajectories $\lambda^{B}_k(X_k)$ can be easily incorporated into the predicted density of Poisson distributed trajectories $\lambda^u_{k|k-1}(X_k)$ that have never been detected. The two different trajectory PMBM filters have different prediction steps; the difference is that whether dead trajectories are still maintained in the set of trajectories. In the update step, a potential new track is initiated for each measurement, and additional hypotheses are created due to data association. The two different trajectory PMBM filters have the same update step. Explicit expressions for how the PMBM parameters (\ref{eq:pmbm_parameters}) are predicted and updated, using the two different problem formulations, can be found in \cite{continuityPMBM}; they are omitted here.

\section{\normalfont Trajectory MBM Filter}
It is shown in \cite{pmbmpoint2} that the MBM RFS of targets is a multi-target conjugate prior if the birth model is a multi-Bernoulli RFS, as in \textbf{Assumption} \ref{assumption:mbbirth}. In this section, we extend this result to RFS of trajectories. Given the sequence of measurements up to time step $k^{\prime}$ and \textbf{Assumption} \ref{assumption:mbbirth} and \ref{assumption:meas}, the density of the set of trajectories at time step $k\in\{k^{\prime},k^{\prime}+1\}$ is given by the MBM of the form
\begin{equation}
    f_{k|k^{\prime}}(\mathbf{X}_k)=\sum_{a\in\mathcal{A}_{k|k^{\prime}}}w^a_{k|k^{\prime}}\sum_{\uplus_{i\in\mathbb{T}_{k|k^{\prime}}}\mathbf{X}_k^i=\mathbf{X}_k}\prod_{i\in\mathbb{T}_{k|k^{\prime}}}f_{k|k^{\prime}}^{i,a^i}(\mathbf{X}^i_k),
    \label{eq:mbm}
\end{equation}
where the MBM RFS $\mathbf{X}_k$ have Bernoulli parameters $r^{i,a^i}_{k|k^{\prime}}$ and $f_{k|k^{\prime}}^{i,a^i}(\cdot)$, cf. (\ref{eq:bernoulli}). A trajectory MBM RFS can be defined by the parameters of the density
\begin{subequations}
    \begin{align}
        &\mathcal{A}_{k|k^{\prime}},\left\{ \Theta_{k|k'}^{a} \right\}_{ a \in \mathcal{A}_{k|k'}},\\
        &\Theta_{k|k'}^{a} = \left\{ \left( w_{k|k'}^{i,a^i} , r_{k|k'}^{i,a^i} , f_{k|k'}^{i,a^i} \right) \right\}_{i \in \mathbb{T}}.
    \end{align}
\end{subequations}

% \begin{equation}
%     \mathcal{A}_{k|k^{\prime}}, \{(w^{i,a^i}_{k|k^{\prime}},r^{i,a^i}_{k|k^{\prime}},f^{i,a^i}_{k|k^{\prime}})\}.
% \end{equation}

\subsection{Structure of the Trajectory MBM Filter}
The structure of the trajectory MBM is similar to the MBM maintained in the trajectory PMBM. The difference lies in how tracks (i.e., Bernoulli components) are initiated. In the trajectory PMBM filter, a new track is initiated for each measurement, whereas in the trajectory MBM filter, a new track is initiated for each Bernoulli component in the multi-Bernoulli birth model, i.e., MBM hypotheses explicitly enumerate potential targets that remain to be detected. Both the trajectory PMBM filter and the trajectory MBM filter can explicitly represent trajectories that remain to be detected. In the PMBM representation, these trajectories are efficiently represented through the trajectory Poisson intensity $\lambda^u_{k|k^{\prime}}(\cdot)$, whereas in the MBM representation, they are split across many single trajectory hypotheses (trajectory Bernoulli RFSs) with empty measurement association history, i.e., $\mathcal{M}^k(i,a^i)=\emptyset$.

% Trajectories that are hypothesized to exist but have never been detected are explicitly represented as a trajectory Poisson RFS in the trajectory PMBM filter, whereas in the trajectory MBM filter, these trajectories are implicitly represented by single trajectory hypotheses (trajectory Bernoulli RFSs) with empty measurement association history, i.e., $\mathcal{M}^k(i,a^i)=\emptyset$.

In each global hypothesis $a\in\mathcal{A}_{k|k}$, each measurement, at each time step, is associated to at most one track, and each track is associated to at most one measurement. Measurements that are not associated to any tracks in a global hypothesis are considered to be clutter under this global hypothesis. Tracks that are not associated to any measurements in a global hypothesis are considered to be misdetected under this global hypothesis.
Under these constraints, the set of global hypotheses at time step $k$ can be expressed as
\begin{multline}
    \mathcal{A}_{k|k^{\prime}}=\bigg\{a=(a^1,...,a^{n_{k|k^{\prime}}})\bigg|\bigcup_{i\in\mathbb{T}_{k|k^{\prime}}}\mathcal{M}^k(i,a^i)\subseteq\mathcal{M}_k,\\\mathcal{M}^k(i,a^i)\cap\mathcal{M}^k(j,a^j)=\emptyset~\forall~i\neq j,~i,j\in\mathbb{T}_{k|k^{\prime}}\bigg\}.
    \label{eq:globalhypothesis_mbm}
\end{multline}
Compared to (\ref{eq:globalhypothesis}), here $\mathcal{M}_k\setminus\bigcup_{i\in\mathbb{T}_{k|k^{\prime}}}\mathcal{M}^k(i,a^i)$ consists of indices of measurements received so far that are clutter under global hypothesis $a\in\mathcal{A}_{k|k^{\prime}}$. This is an important difference from the trajectory PMBM filter, in which the question whether a measurement corresponds to clutter, or to the initialization of a new target trajectory, is captured by the existence probability of the created trajectory Bernoulli RFS.

In the rest of the section, we present the prediction and update steps for recursively computing (\ref{eq:mbm}) for the MBM parameterization. Similar to the trajectory PMBM filter, the two different trajectory MBM filters, based on the set of current trajectories formulation and the set of all trajectories formulation, have the same update step. For compactness, we denote the inner product of two functions $h(\cdot)$ and $g(\cdot)$, as $\langle h;g\rangle = \int h(x)g(x)dx$. 

\subsection{MBM Filtering Recursion}
We first present the prediction steps, respectively, for the two different problem formulations, and then we present the update step.

\subsubsection{Prediction Step for the Set of Current Trajectories}
The prediction step is given in the theorem below.
\begin{theorem}
    Assume that the distribution from the previous time step $f_{k-1|k-1}(\mathbf{X}_{k-1})$ is given by (\ref{eq:mbm}), that the transition model is (\ref{eq:transition_current}), and that the birth model is a trajectory multi-Bernoulli RFS with $n^b_k$ Bernoulli components, each of which has density of the form (\ref{eq:mbirth}). Then the predicted distribution for the next step $f_{k|k-1}(\mathbf{X}_{k})$ is given by (\ref{eq:mbm}), with $n_{k|k-1}=n_{k-1|k-1}+n^b_k$. For tracks continuing from previous time ($i\in\{1,...,n_{k-1|k-1}\}$), the parameters of the MBM are:
    \begin{subequations}
        \begin{align}
            h^i_{k|k-1} &= h^i_{k-1|k-1},\\            w^{i,a^i}_{k|k-1}&=w^{i,a^i}_{k-1|k-1}~\forall~a^i,\\
            r^{i,a^i}_{k|k-1} &= r^{i,a^i}_{k-1|k-1}\langle f^{i,a^i}_{k-1|k-1};P^S_{k-1}\rangle~\forall~a^i,\\
            f^{i,a^i}_{k|k-1}(X) &= \frac{\langle f^{i,a^i}_{k-1|k-1};\pi^cP^S_{k-1}\rangle }{\langle f^{i,a^i}_{k-1|k-1};P^S_{k-1}\rangle}~\forall~a^i.
        \end{align}
        \label{eq:mbmpredictexist}
    \end{subequations}
    \\*
    For new tracks ($i\in\{n_{k-1|k-1}+l\}$, $l\in\{1,...,n^b_k\}$), the parameters of the MBM are:
    \begin{subequations}
        \begin{align}
            h^i_{k|k-1} &= 1,\\
            \mathcal{M}^{k-1}(i,1) &= \emptyset,\\
            w^{i,1}_{k|k-1}&=1,\\
            r^{i,1}_{k|k-1} &= r^{b,l}_{k},\\
            f^{i,1}_{k|k-1}(X) &= f^{B,l}_k(X).
        \end{align}
        \label{eq:newberparas}
    \end{subequations}
    \label{theorem1}
\end{theorem}

\subsubsection{Prediction Step for the Set of All Trajectories}
The prediction step is given in the theorem below.
\begin{theorem}
    Assume that the distribution from the previous time step $f_{k-1|k-1}(\mathbf{X}_{k-1})$ is given by (\ref{eq:mbm}), that the transition model is (\ref{eq:transition_all}), and that the birth model is a trajectory multi-Bernoulli RFS with $n^b_k$ Bernoulli components, each of which has density given by (\ref{eq:mbirth}). Then the predicted distribution for the next step $f_{k|k-1}(\mathbf{X}_{k})$ is given by (\ref{eq:mbm}), with $n_{k|k-1}=n_{k-1|k-1}+n^b_k$. For tracks continuing from previous time ($i\in\{1,...,n_{k-1|k-1}\}$), the parameters of the MBM are:
    \begin{subequations}
        \begin{align}
            h^i_{k|k-1} &= h^i_{k-1|k-1},\\            w^{i,a^i}_{k|k-1}&=w^{i,a^i}_{k-1|k-1}~\forall~a^i,\\
            r^{i,a^i}_{k|k-1} &= r^{i,a^i}_{k-1|k-1}~\forall~a^i,\\
            f^{i,a^i}_{k|k-1}(X) &= \langle f^{i,a^i}_{k-1|k-1};\pi^a \rangle~\forall~a^i.
            \label{eq:mbm_alltra_prediction}
        \end{align}
    \end{subequations}
    For new tracks ($i\in\{n_{k-1|k-1}+l\}$, $l\in\{1,...,n^b_k\}$), the parameters of the MBM are the same as (\ref{eq:newberparas}).
    \label{theorem2}
\end{theorem}

\subsubsection{Update Step}
The update step is given in the theorem below.
\begin{theorem}
    Assume that the predicted distribution $f_{k|k-1}(\mathbf{X}_{k})$ is given by (\ref{eq:mbm}), that the measurement model is (\ref{eq:measurementmodel}), and that the measurement set at time step $k$ is $\mathbf{z}_k=\{z^1_k,...,z^{m_k}_k\}$. Then the updated distribution $f_{k|k}(\mathbf{X}_{k})$ is given by (\ref{eq:mbm}), with $n_{k|k} = n_{k|k-1}$. For each track ($i\in\{1,...,n_{k|k}\}$), a hypothesis is included for each combination of a hypothesis from a previous time and either a misdetection or an update using one of the $m_k$ new measurements, such that the number of hypotheses becomes $h^i_{k|k}=h^i_{k|k-1}(1+m_k)$. For misdetection hypotheses ($i\in\{1,...,n_{k|k}\}, a^i\in\{1,...,h_{k|k-1}\}$), the parameters of the MBM are
    \begin{subequations}
        \begin{align}
            \mathcal{M}^k(i,a^i) &= \mathcal{M}^{k-1}(i,a^i),\\
            w^{i,a^i}_{k|k} &=w^{i,a^i}_{k|k-1}\left(1-r^{i,a^i}_{k|k-1} \left\langle  f^{i,a^i}_{k|k-1};P^D\right\rangle \right),\\
            r^{i,a^i}_{k|k} &=\frac{r^{i,a^i}_{k|k-1}\left\langle f^{i,a^i}_{k|k-1};1-P^D \right\rangle}{1-r^{i,a^i}_{k|k-1} \left\langle  f^{i,a^i}_{k|k-1};P^D\right\rangle},\\
            f^{i,a^i}_{k|k}(X) &= \frac{ (1-P^D_k(X))f^{i,a^i}_{k|k-1}(X) }{\left\langle f^{i,a^i}_{k|k-1};1-P^D \right\rangle}.
        \end{align}
    \end{subequations}
    For hypotheses updating tracks ($i\in\{1,...,n_{k|k}\}$, $a^i = \tilde{a}^i+h^i_{k|k-1}j$, $\tilde{a}^i\in\{1,...,h^i_{k|k-1}\}$, $j\in\{1,...,m_k\}$, i.e., the previous hypothesis $\tilde{a}^i$, updated with measurement $z^j_k$), the parameters are
    \begin{subequations}
        \begin{align}
            \mathcal{M}^k(i,a^i) &= \mathcal{M}^{k-1}(i,\tilde{a}^i)\cup\{(k,j)\},\\
            w^{i,a^i}_{k|k} &=\frac{w^{i,a^i}_{k|k-1}r^{i,\tilde{a}^i}_{k|k-1}\left\langle f^{i,\tilde{a}^i}_{k|k-1};\varphi(z^j_k|\cdot)P^D \right\rangle}{\lambda^{\text{FA}}(z^j_k)},\\
            r^{i,a^i}_{k|k} &= 1,\\
            f^{i,a^i}_{k|k}(X) &= \frac{\varphi(z^j_k|X)P^D_k(X)f^{i,\tilde{a}^i}_{k|k-1}(X)}{\left\langle f^{i,\tilde{a}^i}_{k|k-1};\varphi(z^j_k|\cdot)P^D_k\right\rangle}.
        \end{align}
    \end{subequations}
    \label{theorem3}
\end{theorem}
The derivation here incorporates hypotheses updating every prior hypothesis with every measurement; however, in practical implementations, gating can be used to reduce the computational burden by excluding hypotheses with negligible weights.

\subsection{MBM01 Filtering Recursion}
The trajectory $\text{MBM}_{01}$ filter can be considered as a variant of the trajectory MBM filter, in which existence probabilities of Bernoulli components are either 0 or 1. The $\text{MBM}_{01}$ filtering recursion can be obtained from the MBM filtering recursions by expanding the MBM into its $\text{MBM}_{01}$ equivalent \cite{pmbmpoint2}. The filtering recursions for the trajectory $\text{MBM}_{01}$ filter are given in Appendix \ref{ap:b}.

\subsection{Discussion}
% Four different trajectory filters result from the theorems: a trajectory MBM filter for the set of current trajectories is given by the prediction in Theorem \ref{theorem1} and the update in Theorem \ref{theorem3}; a trajectory MBM filter for the set of all trajectories is given by the prediction in Theorem \ref{theorem2} and the update in Theorem \ref{theorem3}; a trajectory $\text{MBM}_{01}$ filter for the set of current trajectories is given by the prediction in Theorem \ref{theorem4} and the update in Theorem \ref{theorem6}; a trajectory $\text{MBM}_{01}$ filter for the set of all trajectories is given by the prediction in Theorem \ref{theorem5} and the update in Theorem \ref{theorem6}.

All the trajectory filters presented above are track-oriented. For each Bernoulli component in the multi-Bernoulli birth density, a new track is initiated. Compared to the trajectory PMBM filter with Poisson RFS birth, tracks are created in the prediction step but not the update step of trajectory MBM/$\text{MBM}_{01}$ filter. In the trajectory MBM/$\text{MBM}_{01}$ filter for the set of all trajectories, the prediction (\ref{eq:mbm_alltra_prediction}) and (\ref{eq:mbm01_alltra_prediction}) result in additional mixture component in Bernoulli densities $f^{i,a^i}_{k|k^{\prime}}(X_k)$, which are of the form
\begin{equation}
    p(X) = \sum_jw^jp^j(x_{\beta:\varepsilon}|\beta,\varepsilon)\Delta_{e^j}(\varepsilon)\Delta_{b^j}(\beta),
    \label{eq:trajectorymixture}
\end{equation}
where each mixture component is characterized by a weight $w^j$, a distinct birth time $b^j$, a distinct most recent time $e^j$ where $b^j\leq e^j$ for all $j$\footnote{Neither the birth time $\beta$ nor the most recent time $\varepsilon$ is deterministic.}, and a state sequence density $p^j(\cdot)$. This type of state density facilitates simple representations for the state sequence $x_{\beta:\varepsilon}$ (either the state of a trajectory that is still present, or the state of a dead trajectory), conditioned on $\beta$ and $\varepsilon$.

The prediction steps, given by Theorem \ref{theorem4} and Theorem \ref{theorem5}, in the trajectory $\text{MBM}_{01}$ filter, create more single trajectory hypotheses than the prediction steps, given by Theorem \ref{theorem1} and Theorem \ref{theorem2}, in the trajectory MBM filter; this is a direct result of restricting the existence probability of Bernoulli components to either 0 or 1. The existence probability of trajectory Bernoulli RFS $r$ has different meanings in the four different trajectory filters: in the trajectory MBM filter for the set of current trajectories, $r$ is the probability that the trajectory exists at the current time and has not ended yet; in the trajectory MBM filter for the set of all trajectories, $r$ represents the probability that the trajectory existed at any time before including the current time; in the trajectory $\text{MBM}_{01}$ filter for the set of current trajectories, $r$ indicates whether the trajectory exists at the current time and has not ended yet; in the trajectory $\text{MBM}_{01}$ filter for the set of all trajectories, $r$ indicates whether the trajectory existed at any time before and including the current time.

We remark that the labelled trajectory MBM and  $\text{MBM}_{01}$ filters, which are defined over the set of labelled trajectories, can be obtained by augmenting label to single target state $x$ \cite[Sec. IV-A]{trackingbasedontrajectories}. This does not affect the filtering recursion or the information in the computed posterior, compared to MBM and $\text{MBM}_{01}$. Therefore, the corresponding multi-scan implementations in Section \ref{para:implementation} are analogous.

\section{\normalfont Implementation of Multi-Scan Trajectory Filters}
\label{para:implementation}

In this section, we present efficient multi-scan implementations of the above trajectory filters.

% \subsection{State sequence density}

% An implementation of the PMBM trajectory filter

% If the reconstruction is not feasible, we can instead use a greedy heuristic method \cite{greedyheuristics} to find a feasible primal solution.

\subsection{Hypothesis Reduction}

% In the trajectory filter, the posterior over the set of detected trajectories is an MBM, i.e., a weighted mixture of global hypotheses. In the update step, each possible data association creates an updated global hypothesis, such that the number of multi-Bernoullis in the posterior MBM density increases exponentially over time. Finding a feasible approach to reduce the number of global hypotheses after the update step is, therefore, essential for designing a computationally tractable trajectory filter. 

The hypothesis reduction techniques for the trajectory PMBM, MBM and $\text{MBM}_{01}$ are quite similar so we first explain the general formulation and then highlight the differences. As a first step, we identify the most probable global hypothesis, from which estimates of trajectories are also typically extracted. Conditioning on the most likely global hypothesis, we make use of track-oriented $N$-scan pruning \cite{blackman2004multiple}, a conventional hypothesis reduction technique used in TOMHT, to prune global hypotheses with negligible weights. 

We note that hypothesis reduction is not complicated by the fact that we are working with symmetric (unlabelled) distributions. Specifically, in (20), the quantities stored are the weight of hypothesis $a$, i.e., $w^a_{k|k\prime}$, and the hypothesis-conditioned trajectory distributions $f^{i,a^i}_{k|k\prime}(\mathbf{X}_k^i)$ for each target. Symmetry is ensured by the sum over $\uplus_{i\in\mathbb{T}_{k|k\prime}}\mathbf{X}_k^i=\mathbf{X}_k$; this sum is implicit, and terms never need to be explicitly represented. Therefore, hypothesis reduction achieved by either setting $w^a_{k|k\prime}=0$ for some subset of hypotheses (and re-normalising the weights of remaining hypotheses to sum to 1), or by removing a subset of multi-Bernoulli components $f^{i,a^i}_{k|k\prime}(\mathbf{X}_k^i)$ for some hypotheses, always results in valid symmetric distributions. Likewise, if the existence probability of a Bernoulli component is close to zero in all the considered global hypotheses, pruning is equivalent to setting this existence probability equal to zero, which does not affect the symmetry of the posterior.

% Pruning consists of approximating some of the weights $w^a_{k|k^{\prime}}$ that represent the filtering (multi-trajectory) density (\ref{eq:pmbm}), (\ref{eq:mbm}) as zero, followed by a normalization of the resulting weight.

% After pruning, we obtain another PMBM, MBM or $\text{MBM}_{01}$ distribution, which is symmetric by construction.

Given the most likely global hypothesis $a^{*}$ at current time step $k$, we trace the single trajectory hypotheses included in $a^{*}$ back to their local hypotheses at time step $k-N$. The assumption behind the $N$-scan pruning method is that the data association ambiguity is resolved before scan $k-N$ \cite{blackman2004multiple}. In other words, global hypotheses that do not coincide with $a^{*}$ up until and including time step $k-N+1$ are assumed to have negligible weights; these global hypotheses can then be pruned. In addition, tracks (local hypotheses trees) which, after pruning, have a single non-existence local hypothesis, i.e., $r=0$, can be pruned. In what follows, we show that the most likely global hypothesis $a^{*}$ can be obtained as the solution of a multi-frame assignment problem.

\subsection{Data Association Modeling and Problem Formulation}
% The global hypothesis weights $\mathcal{W}^A$ are related to the single target hypothesis weights via the expression \cite{pmbmpoint}
% \begin{equation}
%     \mathcal{W}^A\propto\prod_{i\in\mathbb{X}}w^{\mathbf{a}^i},
%     \label{eq:globalhypothesisweight}
% \end{equation}
% where the proportionality denotes that normalization is required to ensure that $\sum_{A\in\mathfrak{A}}\mathcal{W}^A=1$.

As indicated in the previous section, the posterior global hypothesis probability $w^a_{k|k}$ is proportional to the product of the weights of different single trajectory hypotheses $w^{i,a^i}_{k|k}$, one from each track:
\begin{equation}
    w^a_{k|k}\propto\prod_{i\in\mathbb{T}_{k|k}}w^{i,a^i}_{k|k},
    \label{eq:globalhypothesisweight}
\end{equation}
where the proportionality denotes that normalization is required to ensure that $\sum_{a\in\mathcal{A}_{k|k}}w^{a}_{k|k}=1$. Omitting time indices and introducing the notation $c^a=-\log (w^a)$ and $c^{i,a^i}=-\log (w^{i,a^i})$, yields
\begin{equation}
    c^a = \sum_{i\in\mathbb{T}}c^{i,a^i} + C,
\end{equation}
where $C$ is the logarithm of the normalization constant in (\ref{eq:globalhypothesisweight}). The most likely global hypothesis is the collection of single trajectory hypotheses that minimizes the total cost, i.e., 
\begin{equation}
    a^{*}=\underset{(a^i)\in\mathcal{A}}{\arg\min}\sum_{i\in\mathbb{T}}c^{i,a^i}.
    \label{eq:globalminimize}
\end{equation}

Let $\mathcal{H}^i$ denote the set of single trajectory hypotheses for the $i$th track, and let $\mathbb{M}_{\tau}$ denote the set of measurement indices at time step ${\tau}$. Further, let $\rho^{i,a^i}\in\{0,1\}$ be a binary indicator variable, indicating whether single trajectory hypothesis $a^i$ in the $i$th track is included in a global hypothesis or not, and let 
\begin{equation}
    \boldsymbol{\rho}=\left\{\rho^{i,a^i}\in\{0,1\} \Big| a^i\in\mathcal{H}^i~\forall ~i\in\mathbb{T}\right\}
\end{equation}
be the set of all binary indicator variables. The minimization problem (\ref{eq:globalminimize}) can be further posed as a multi-frame assignment problem by decomposing the constraint $(a^i)\in\mathcal{A}$ into a set of smaller constraints \cite[Section III]{dualdecomposition}, in the form of
\begin{equation}
    \underset{\boldsymbol{\rho} \in \bigcap_{\tau=0}^{k} \mathcal{P}^{\tau} }{\arg\min}\sum_{i\in\mathbb{T}}\sum_{a^i\in\mathcal{H}^i}c^{i,a^i}\rho^{i,a^i},
    \label{eq:costfunction}
\end{equation}
with the constraints sets denoted as
\begin{subequations}
    \begin{align}
        \mathcal{P}^0 &= \left \{\boldsymbol{\rho}\bigg|\sum_{a^i\in\mathcal{H}^i}\rho^{i,a^i}=1,~\forall ~i\in\mathbb{T}\right\},\label{eq:constraint1}\\
        \mathcal{P}^{\tau} &=\Bigg \{\boldsymbol{\rho}\Bigg|\sum_{i\in\mathbb{T}}\sum_{\substack{a^i\in\mathcal{H}^i:\\(\tau,j)\in\mathcal{M}(i,a^i)}}\rho^{i,a^i}\leq 1,~\forall~ j\in\mathbb{M}_{\tau}\Bigg\},
    \label{eq:constraint2}
    \end{align}
\end{subequations}
% \begin{equation}
%     \mathcal{P}^0 = \left \{\boldsymbol{\rho}\bigg|\sum_{a^i\in\mathcal{H}^i}\rho^{i,a^i}=1,~\forall ~i\in\mathbb{T}\right\},
%     \label{eq:constraint1}
% \end{equation}
% \begin{equation}
%     \mathcal{P}^k =\Bigg \{\boldsymbol{\rho}\Bigg|\sum_{i\in\mathbb{T}}\sum_{\substack{a^i\in\mathcal{H}^i:\\(k,j)\in\mathcal{M}(i,a^i)}}\rho^{i,a^i}\leq 1,~\forall~ j\in\mathbb{M}_k\Bigg\},
%     \label{eq:constraint2}
% \end{equation}
where $k$ is the current time step and $\tau=1,...,k$. The first constraint (\ref{eq:constraint1}) enforces that each global hypothesis should include one and only one single trajectory hypothesis from each track. The set of $k$ constraints (\ref{eq:constraint2}) differs in the trajectory PMBM filter and the trajectory MBM/$\text{MBM}_{01}$ filter. In the trajectory PMBM filter, each measurement from each time should be associated to exactly one track, i.e., the $\leq$ sign becomes an $=$ sign in (\ref{eq:constraint2}), whereas in the trajectory MBM/$\text{MBM}_{01}$ filter, each measurement from each time should be associated to at most one track, which explains the $\leq$ sign.

\subsection{Multi-Frame Assignment via Dual Decomposition}

% \subsection{Previous approach}
% For each predicted global hypothesis there are multiple possible data associations, each of which will result in a MB component in the updated MB mixture. In the previous work \cite{pmbmpoint2}, the Murty's algorithm \cite{murty} is first used to select the $k$-best new global hypotheses with the highest $k$ weights for each predicted global hypotheses. Then, global hypotheses are further truncated by only keeping those with high weights. The best global hypothesis can be either selected as the one with the maximum weight, or the one with the MAP cardinality \cite{pmbmpoint2}.

% \subsubsection{Dual decomposition}
The multi-dimensional assignment problem (\ref{eq:costfunction}) is NP-hard for two or more scans of measurements. An effective approach to solving this problem is Lagrangian relaxation; this technique has been widely used to solve the multi-scan data association problem in TOMHT algorithms, see, e.g., \cite{lagrange1,lagrange2}. In this work, we focus on the dual decomposition formulation \cite{mrfdualdecomposition}, i.e., a special case of Lagrangian relaxation, whose competitive performance, compared to traditional approaches \cite{lagrange1,lagrange2}, in solving the multi-frame assignment problem has been demonstrated in \cite{dualdecomposition}.

% Note that, in a practical implementation, a sliding window over the last $S$-scan should be used to further reduce the computational complexity.
\subsubsection{Decomposition of the Lagrangian Dual}
We follow similar implementation steps as in \cite{dualdecomposition}. The original (primal) problem (\ref{eq:costfunction}) is separated into $k$ subproblems, one for each time step, and for each subproblem a binary variable is used. The subproblem solutions
\begin{equation}
    \boldsymbol{\rho}_{\tau} =\{\rho_{\tau}^{i,a^i}\in\{0,1\}|a^i\in\mathcal{H}^i~\forall~i\in\mathbb{T}\},
\end{equation} 
must be equal for all ${\tau}$; this is enforced through Lagrange multipliers that are incorporated into the subproblems acting as penalty weights. The ${\tau}$th subproblem can be written as \cite{dualdecomposition}
\begin{multline}
    \underset{{\boldsymbol{\rho}_{\tau}\in\mathcal{P}^0\cap\mathcal{P}^{\tau}}}{\arg\min}\sum_{i\in\mathbb{T}}\sum_{a^i\in\mathcal{H}^i}\bigg(\frac{c^{i,a^i}}{k}+\delta_{\tau}^{i,a^i}\bigg)\rho_{\tau}^{i,a^i} \\ \triangleq \underset{{\boldsymbol{\rho}_{\tau}\in\mathcal{P}^0\cap\mathcal{P}^{\tau}}}{\arg\min}\mathcal{S}(\boldsymbol{\rho}_{\tau},\boldsymbol{\delta}_{\tau}),
    \label{eq:subproblem}
\end{multline}
where the Lagrange multipliers used for the ${\tau}$th subproblem are denoted by
\begin{equation}
    \boldsymbol{\delta}_{\tau}=\{\delta_{\tau}^{i,a^i}|a^i\in\mathcal{H}^i~\forall~i\in\mathbb{T}\},
\end{equation}
and the division by $k$ in (\ref{eq:subproblem}) comes from the fact that the summation of the objectives that each subproblem tries to minimize should be equal to the objective of the original problem.
The Lagrange multipliers $\delta_{\tau}^{i,a^i} \in \mathbb{R}$ have the constraint that, for each single trajectory hypothesis, they must add up to zero over different subproblems  \cite{mrfdualdecomposition}. Thus, the set of Lagrange multipliers has the form
\begin{equation}
    \Lambda=\left\{\boldsymbol{\delta}_{\tau}\Bigg|\sum_{{{\tau}}=1}^{k}\delta_{\tau}^{i,a^i}=0,~\forall~a^i\in\mathcal{H}^i~\forall~i\in\mathbb{T}\right\}.
\end{equation}

\subsubsection{Subproblem Solving}
After eliminating all the constraints sets except two, i.e., $\mathcal{P}^0$ and $\mathcal{P}^{{\tau}}$, we obtain a 2-D assignment problem (\ref{eq:subproblem}). The objective of the ${\tau}$th assignment problem (\ref{eq:subproblem}) is to 
associate each measurement received at time step ${\tau}\leq k$, i.e., $j\in\mathbb{M}_{\tau}$, either to an existing track or a new track\footnote{In the trajectory MBM/$\text{MBM}_{01}$, ``dummy'' tracks are created to represent clutter.} at the current time step $k$, i.e., $i\in\mathbb{T}_{k}$, such that the total assignment cost is minimized.

% Note that the formulation of this assignment problem is different in the trajectory PMBM filter and the trajectory MBM/$\text{MBM}_{01}$ filter. In the trajectory PMBM filter, the objective of the $k$th assignment problem is to 
% associate each measurement received at time $k\leq\tau$, i.e., $j\in\mathbb{M}_k$, either to an existing track or a new track at current time $\tau$, i.e., $i\in\mathbb{T}_{\tau}$, such that the total assignment cost is minimized. As a comparision, in the trajectory MBM filter, the corresponding objective is to associate each measurement received at time $k\leq\tau$, either to a track or clutter. Problems of this type can be solved in polynomial time using a modified auction algorithm \cite[Chapter VII]{bar1990multitarget}. Note that the standard assignment formulation involves rows and columns of the assignment matrix summing to one. To explicitly represent the assignment of a measurement to clutter in the trajectory MBM filter, the involved cost matrix needs to be reformulated by introducing dummy variables.

For a track that is created after time step ${\tau}$, no measurement from time step ${\tau}$ should be assigned to it; therefore, the measurement-to-track assignment cost is infinity. For a track that existed before and up to time step ${\tau}$, i.e., $i\in\mathbb{T}_{\tau}$, if measurement $z^j_{\tau}$ was not associated to this track, let the measurement-to-track assignment cost be infinity; if otherwise, let the cost first be the minimum cost of the single trajectory hypothesis in this track that was updated by $z^j_{\tau}$ \cite[Chapter VII, Equation (7.24)]{bar1990multitarget}, i.e.,
\begin{equation}
\min\sum_{\substack{a^i\in\mathcal{H}^i:\\({\tau},j)\in\mathcal{M}(i,a^i)}}\left(\frac{c^{i,a^i}}{k}+\delta_{\tau}^{i,a^i}\right).
\label{eq:minisinglecost}
\end{equation}

In order to keep the cost of a hypothesis that does not assign a measurement to a track the same for an existing track and a new track (trajectory PMBM filter) or clutter (trajectory MBM filter), the cost (\ref{eq:minisinglecost}) should then have subtracted from it by the minimum cost of hypotheses that this track is not updated by any of the measurements at time step ${\tau}$, i.e.,
\begin{equation}
    \min\sum_{\substack{a^i\in\mathcal{H}^i:\\({\tau},j)\notin\mathcal{M}(i,a^i),\forall j\in\mathbb{M}_{\tau}}}\left(\frac{c^{i,a^i}}{k}+\delta_{\tau}^{i,a^i}\right).
\end{equation}
Note that, in the context of Lagrangian relaxation, the costs of single trajectory hypotheses refer to the costs that are penalized by the Lagrangian multipliers.

% Note that assignment problems with non-square matrices can be reformulated with square matrices by introducing dummy variables. 
After solving the 2-D assignment problem, we can obtain the associations for each measurement at time step ${\tau}$. For tracks not being associated to any measurements at time step ${\tau}$, if the track is created before and up to time step ${\tau}$, i.e., $i\in\mathbb{T}_{\tau}$, the single trajectory hypothesis
\begin{equation}
\argmin_{a^i}\sum_{\substack{a^i\in\mathcal{H}^i:\\({\tau},j)\notin\mathcal{M}(i,a^i),\forall j\in\mathbb{M}_{\tau}}}\left(\frac{c^{i,a^i}}{k}+\delta_{\tau}^{i,a^i}\right)
\end{equation}
is included in the most likely global hypothesis; if otherwise, i.e., $i\in\mathbb{T}_{k}\setminus\mathbb{T}_{\tau}$, we can choose the single trajectory hypothesis
\begin{equation}
\argmin_{a^i}\sum_{a^i\in\mathcal{H}^i}\left(\frac{c^{i,a^i}}{k}+\delta_{\tau}^{i,a^i}\right)
\end{equation}
to be included in the most likely global hypothesis.

\subsubsection{Subgradient Updates}
The objective of Lagrange relaxation is to find the tightest lower bound of the summation of the cost of each subproblem (\ref{eq:subproblem}). The dual problem can be expressed as \cite{dualdecomposition}
\begin{equation}
    \underset{\{\boldsymbol{\delta}_{\tau}\}\in\Lambda}{\arg\max}\bigg(\sum_{{\tau}=1}^{k}\min_{\boldsymbol{\rho}_{\tau}\in\mathcal{P}^0\cap\mathcal{P}^{\tau}}\mathcal{S}(\boldsymbol{\rho}_{\tau},\boldsymbol{\delta}_{\tau})\bigg),
    \label{eq:dual}
\end{equation}
where the maximum can be found using subgradient methods \cite{subgradient}. The Lagrange multipliers $\{\boldsymbol{\delta}_{\tau}\}$ are updated  using
\begin{equation}
    \delta_{\tau}^{i,a^i} = \delta_{\tau}^{i,a^i} + \alpha_t\cdot g_{\tau}^{i,a^i},
\end{equation}
where $g_{\tau}^{i,a^i}$ is the projected subgradient that can be calculated as
\begin{equation}
    g_{\tau}^{i,a^i} = \rho_{\tau}^{i,a^i}-\frac{1}{k}\sum_{{\tau}^{\prime}=1}^{k}\rho_{{\tau}^{\prime}}^{i,a^i},
    \label{eq:constraintLagrange}
\end{equation}
and $\alpha_t$ is the step size at iteration $t$. There are many rules to set the step size, see \cite{mrfdualdecomposition}. In this work, we choose to use the same setting as in \cite{dualdecomposition}, which has the form
\begin{equation}
    \alpha_t = \frac{C^{\text{BP}}_t-C^{D}_t}{\|\{g_{\tau}\}\|^2},
\end{equation}
where $C^{\text{BP}}_t$ is the best (minimum) feasible primal cost so far obtained, $C^{D}_t$ is the dual cost calculated at iteration $t$ from (\ref{eq:dual}), and $\{g_{\tau}\}$ denotes the concatenation of all the projected subgradients $g_{\tau}^{i,a^i}$. The optimal solution is assumed to be attained when the relative gap between the primal cost and the dual cost $(C^{\text{BP}}_t-C^{D}_t)/C^{\text{BP}}_t$ is less than a specified threshold, e.g., 0.01. \cite{mrfdualdecomposition}.

Each subproblem solution will, in general, be infeasible with respect to the primal problem (\ref{eq:costfunction}); nevertheless, subproblem solutions will usually be nearly feasible since large constraints violations were penalized \cite{mrfdualdecomposition}. Hence, feasible solutions $\boldsymbol{\rho}$ can be obtained by correcting the minor conflicting binary elements on which subproblem solutions $\boldsymbol{\rho}_{\tau}$ disagree. Tracks for which we have not yet selected which single trajectory hypothesis to be included in the most likely global hypothesis, we use the branch and bound technique \cite{lawler1966branch} to reconstruct the best feasible solution at each iteration of the Lagrange relaxation. Note that there are many other ways to recover a feasible primal solution from subproblem solutions, see \cite{mrfdualdecomposition}.

\subsection{Discussion}

The objective of solving the multi-frame assignment problem is to know which Bernoulli components are included in the multi-Bernoulli with the highest weight. Because the data association ambiguity is assumed to be resolved before time step $k-N$, obtaining the most likely global hypothesis at time step $k$, which explains the origin of each measurement from time step $k-N$ to current time step $k$, requires the solution of a $N+2$ dimensional assignment problem \cite{blackman2004multiple}.

The computational complexity of filters can be further reduced by limiting the number of single target/trajectory hypotheses, see \cite{pmbmpoint2,continuityPMBM}. As for the multi-scan trajectory PMBM and MBM filters, pruning single trajectory hypotheses with small existence probabilities besides $N$-scan pruning might sometimes harm the solvability of the multi-frame assignment problem, since the problem is formulated using the measurement assignment information contained in single trajectory hypotheses. Instead, we can choose single trajectory hypotheses $a^i\in\mathcal{H}^i,~\forall~ i\in\mathbb{T}$ with small Bernoulli existence probability $r$ at current time step to be updated only by misdetection at next time step. Then single trajectory hypotheses with several consecutive misdetections can be pruned using $N$-scan pruning. Also, to limit the number of mixture components in the trajectory Poisson RFS, components with negligible weights can be pruned.

\section{\normalfont Efficient Fixed-Lag Smoothing}
Multi-target filters based on sets of trajectories are able to estimate the full state sequence instead of appending the sequence of estimates at each time step. This is possible since the posterior density contains full trajectory information.  The posterior density over the set of trajectories can be computed either off-line by applying fixed-interval smoothing, or recursively as new measurements arrive by performing smoothing-while-filtering. Examples of the latter case include the Gaussian mixture trajectory (cardinalized) probability hypothesis density filter proposed in \cite{garcia2018trajectory,garcia2018trajectorycphd} and the trajectory $\text{MBM}_{01}$ filter proposed in \cite{trackingbasedontrajectories} that use an accumulated state density representation \cite{koch2011accumulated}, and the trajectory PMBM filter proposed in \cite{continuityPMBM} that uses an information form \cite{eustice2006exactly}, to represent the joint state density. 

\begin{figure}[!t]
    \centering
    \subfloat[][Scenario 1]{\includegraphics[width=0.23\textwidth]{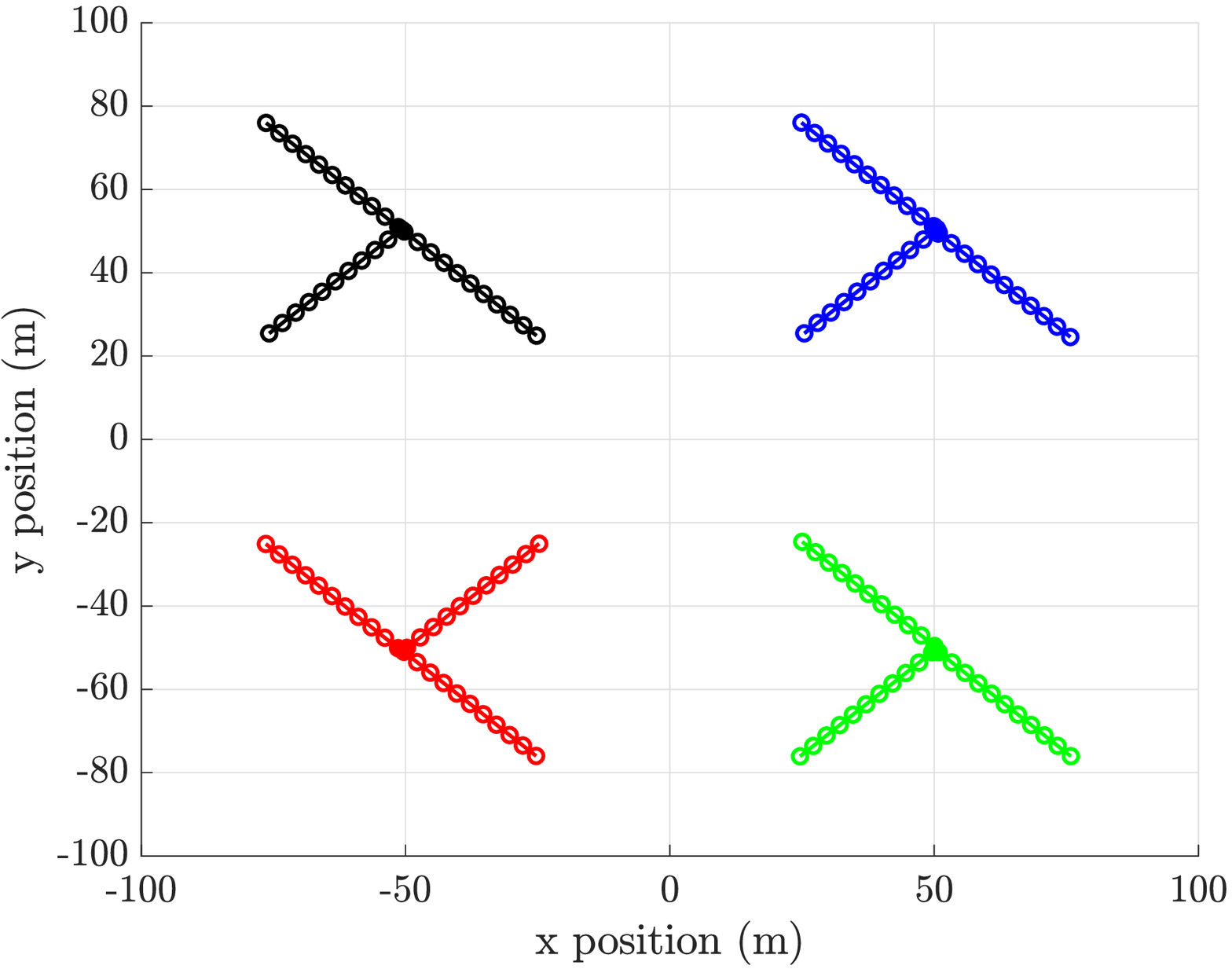}
        \label{fig:scenario1}}
    \subfloat[][Scenario 2]{\includegraphics[width=0.23\textwidth]{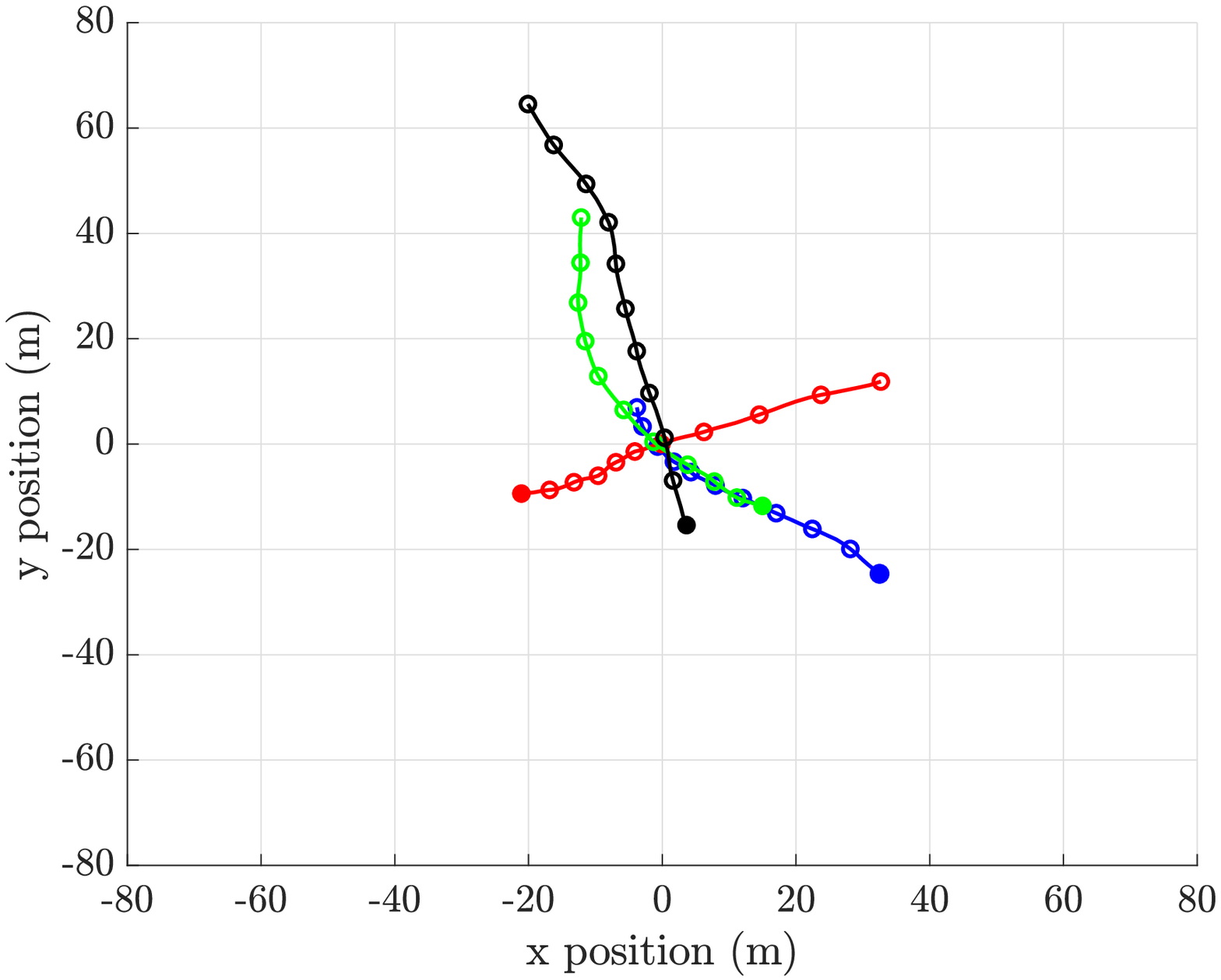}
        \label{fig:scenario2}}
    \caption{True target trajectories for 81 time steps. In both scenarios, targets are born at times \{1,11,21,31\} and are dead at times \{51,61,71,81\}. Targets positions every 6 time steps are marked with a circle, and their initial positions with a filled circle. In Scenario 1, there are twelve targets born at four different locations. In Scenario 2, targets move in close proximity around the mid-point.}
    \label{fig:groundtruth}
\end{figure}

As time progresses, the lengths of the trajectories increase. Eventually, the length may be such that it is computationally beneficial to perform approximate smoothing-while-filtering. An $L$-scan implementation is proposed in \cite{garcia2018trajectory,trackingbasedontrajectories} that propagates the joint density of the states of the last $L$ time steps and independent densities for the previous states for each trajectory. Still, from the perspective of $N$-scan pruning, a lot of unnecessary calculations might be spent on obtaining the smoothed posterior density for each single trajectory hypothesis. More specifically, when the data association ambiguity is high (e.g., targets move in proximity), we might have hundreds or even thousands of single trajectory hypotheses, and at each time instance we only need to compute the posterior trajectory mean for those that are included in the most likely global hypothesis. However, note that the prediction and update of the hypotheses weights are the same as in the implementation using smoothing-while-filtering, e.g., \cite{continuityPMBM}.

% Naturally, it would be more computationally efficient if we only compute the posterior trajectory mean for the most likely global hypothesis. However, a challenge with this approach is that the posterior mean at time $\tau-L$ and $\tau-L-1$ has not generally been computed yet for the most likely global hypothesis, since the trajectory states extracted from the most likely global hypothesis at time $k$ are not necessarily the continuations of the trajectory states extracted at time $k-1$. Instead, we might have to perform smoothing beyond the last $L$ time steps. 

We propose an efficient fixed-lag smoothing implementation of multi-scan trajectory filters that solves the above mentioned problem by combining the $L$-scan trajectory density approximation with $N$-scan pruning. After $N$-scan pruning, single trajectory hypotheses in the same track share the same measurement association history at all times up to time step $k-N$. Then we can apply $(N+L)$-scan density approximation, such that all single trajectory hypotheses in the same track share the same posterior trajectory density up until time step $k-N-L$. It is therefore sufficient to perform fixed-lag smoothing for $N+L$ steps for the most likely global hypothesis, and then store the parameters of the smoothed target state densities at time step $k-N-L+1$ before proceeding. Following this approach, the extracted posterior trajectory mean from the most likely global hypothesis at time step $k+1$ consists of the newly computed smoothed estimates for the last $N+L$ steps and the prestored smoothed estimates at all times up to $k-N-L+1$.

\section{\normalfont Simulations}
% A comparison study of multi-target filters based on multi-target conjugate prior for set of targets can be found in \cite{performanceevaluation}; moreover, a performance evaluation of the multi-scan trajectory PMBM filter, the PMBM filter for set of targets \cite{pmbmpoint2} and the $\delta$-GLMB filter \cite{gibbsglmb} is given in the prelimilary work \cite{xia2018implementation} of this paper. 
In this section we show simulation results that compare five different filters\footnote{The TOMHT implementation developed in \cite{dualdecomposition} can be considered as a special case of the multi-scan trajectory PMBM filter for sets of current trajectories where
the trajectory estimates compose of target state estimates that are extracted from the marginal densities over the current set of targets. Therefore, we choose not to include the TOMHT implementation
in \cite{dualdecomposition} in the simulation results.}: 
\begin{enumerate}
    \item multi-scan trajectory PMBM filter\footnote{\texttt{MATLAB} code of the multi-scan trajectory PMBM, MBM and $\text{MBM}_{01}$ filters is available at https://github.com/yuhsuansia/Multi-scan-trajectory-PMBM-filter. \label{ft7}},
    \item multi-scan trajectory MBM filter\footnotemark[7],
    \item multi-scan trajectory $\text{MBM}_{01}$ filter\footnotemark[7],
    \item fast implementation of the $\delta$-GLMB filter using Gibbs sampling\footnote{We use the code that Profs Ba-Ngu Vo and Ba-Tuong Vo share online: http://ba-tuong.vo-au.com/codes.html. The authors thank them for providing the code. \label{ft2}} \cite{gibbsglmb},
    \item fast implementation of the LMB filter using Gibbs sampling\footnotemark[8] \cite{reuter2017fast}.
\end{enumerate}
For all the trajectory filters, we consider the set of all trajectories problem formulation.

% In this section, we show simulation results that compare the proposed PMBM trajectory filter with the PMBM target filter \cite{pmbmpoint2}, and the $\delta$-GLMB filter with joint prediction and update steps \cite{gibbsglmb}. For the $\delta$-GLMB filter and the PMBM target filter, the $M$ most likely data associations are approximately found using the Gibbs sampling solution proposed in \cite{gibbsglmb}. In the implementation, all the codes are written in \texttt{MATLAB}, except the Gibbs sampling and the auction algorithm, which are written in \texttt{C++}. 

% \footnote{In this work, we focus on the implementation of the $\delta$-GLMB filter using Murty's algorithm. An alternative is to use a Gibbs sampling based method to truncate the GLMB filtering density \cite{gibbsglmb}. The same applies to the existing implementation of the PMBM filter \cite{pmbmpoint2}.}
\subsection{Parameter Setup}

A two-dimensional Cartesian coordinate system is used to define measurement and target kinematic parameters. The kinematic target state is a vector of position and velocity $x_k=[p_{x,k},v_{x,k},p_{y,k},v_{y,k}]^T$. A single measurement is a vector of position $z_k=[z_{x,k},z_{y,k}]^T$. Targets follow a linear Gaussian constant velocity model $\pi_{k|k-1}(x_k|x_{k-1}) = \mathcal{N}(x_k;F_kx_{k-1},Q_k)$, with parameters
\begin{equation*}
    F_k=I_2 \otimes \begin{bmatrix}
        1 & T\\
        0 & 1
    \end{bmatrix}, \quad Q_k =  0.01I_2\otimes\begin{bmatrix}
        T^3/3 & T^2/2\\
        T^2/2 & T
    \end{bmatrix},
\end{equation*}
where $\otimes$ is the Kronecker product, $I_m$ is an identity matrix of size $m\times m$, and $T=1$. The linear Gaussian measurement likelihood model has density $f(z_k|x_k) = \mathcal{N}(z_k;H_kx_k,R_k)$, with parameters $H_k = I_2\otimes[1,0]$ and $R_k = I_2$. 

The filters consider that there are no targets at time step 0. For multi-scan trajectory filters, we use $N$-scan pruning ($N=3$) to remove unlikely global hypotheses. In addition to filtering, we also perform fixed-lag smoothing for the lastest four steps. Both filtering and smoothing performance are analyzed. For the trajectory PMBM filter and the trajectory MBM filter, Bernoulli components with existence probability smaller than $10^{-3}$ are not updated by measurements, see Section V-D. For the trajectory PMBM filter, we remove mixture components in the trajectory Poisson RFS with weights smaller than $10^{-3}$. For the $\delta$-GLMB filter, the cap on the number of components  $H^{\textrm{max}}=2000$. Ellipsoidal gating is used in all the compared filters; the gating size in probability is 0.999. 

% \subsection{Parameter setting}
We consider two different scenarios with true trajectories shown in Figure \ref{fig:groundtruth}. In Scenario 1, targets are well-spaced, and there is at most one target born at the same location per scan. In Scenario 2, for each trajectory, we initiate the midpoint from a Gaussian with mean $[0,0,0,0]^T$ and covariance matrix $I_4$, and the rest of the trajectory is generated by running forward and backwards dynamics. This scenario is challenging due to the fact that all the four targets move in close proximity around the mid-point. In the simulation, we consider constant target survival probability $P^S=0.99$, constant target detection probability $P^D=0.9$, and Poisson clutter uniform in the region of interest with rate $\lambda^{\text{FA}}=10$.

For the trajectory PMBM filter, the Poisson birth intensity has the form $\lambda^b_k(x_k) =\sum_{l} 0.05\mathcal{N}(x;\bar{x}^{b,l}_k,P^{b,l}_k)$. For the trajectory MBM filter, the trajectory $\text{MBM}_{01}$ filter, the $\delta$-GLMB filter and the LMB filter, the $l$th Bernoulli component in the multi-Bernoulli birth has existence probability $r^{b,l}_k=0.05$ and single target state density $\mathcal{N}(x;\bar{x}^{b,l}_k,P^{b,l}_k)$. In Scenario 1, we set $\bar{x}_k^{b,1}=[50,0,50,0]^T$, $\bar{x}_k^{b,2}=[50,0,-50,0]^T$, $\bar{x}_k^{b,3}=[-50,0,50,0]^T$, $\bar{x}_k^{b,4}=[-50,0,-50,0]^T$ and $P^{b,l}_k=\texttt{diag}([4,1,4,1])$. In Scenario 2, we set $\bar{x}_k^{b,1}=[0,0,0,0]^T$ and $P^{b,1}_k=\texttt{diag}([100^2,1,100^2,1])$, which covers the region of interest. It should be noted that in the multi-Bernoulli and Poisson birth model have the same intensity (probability hypothesis density) \cite[Eq. (4.129)]{rfs}. This implies that birth models are as close as possible in the sense of Kullback-Leibler divergence.

\subsection{Performance Evaluation}

For all the three multi-scan trajectory filters we estimate the full trajectories directly from the most likely global hypothesis. For the trajectory filters, we choose the most likely cardinality estimate $n^{\star}$ from the multi-Bernoulli of the most likely global hypothesis. We then report trajectory estimates from the $n^{\star}$ Bernoulli components with the highest existence probabilities. Given a Bernoulli state density (\ref{eq:trajectorymixture}), an estimate of the trajectory is obtained by selecting the most probable mixture component $j^*=\arg\max_jw^j_{k|k^{\prime}}$ and reporting its mean value \cite{continuityPMBM}. For the $\delta$-GLMB filter and the LMB filter, we first obtain the maximum a posteriori estimate of the cardinality. We then find the global hypothesis with this cardinality with highest weight and report the mean of the targets in this hypothesis \cite{glmbpoint}. Trajectories are formed by connecting target estimates with the same label.

To evaluate the filtering performance, we used the generalized optimal sub-pattern assignment (GOSPA) metric \cite{gospa}, which can be decomposed into localization cost, missed target cost, and false target cost. The GOSPA metric is applied to the set of current target states at each time step. To evaluate the tracking performance, the trajectory metric in \cite{trajectorymetric} based on linear programming (LP) was used, which can be decomposed into localization cost, missed target cost, false target cost, and track switch cost.

\begin{figure}[!t]
    \centering
    \includegraphics[width=0.48\textwidth]{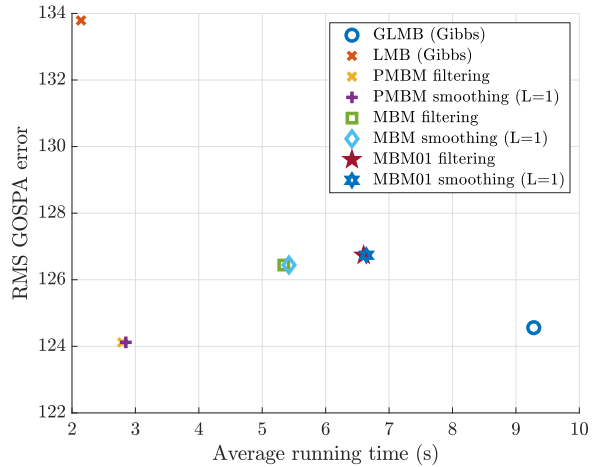}
    \caption{Performance comparison among the $\delta$-GLMB (Gibbs) filter, the LMB (Gibbs) filter, the trajectory PMBM filter, the trajectory MBM filter and the trajectory $\text{MBM}_{01}$ filter in Scenario 2: RMS GOSPA error versus average running time.}
    \label{fig:time_vs_error}
\end{figure}

\begin{table*}[!t]
\centering
\caption{Simulation results for Scenario 1: RMS GOSPA/LP trajectory metric errors and average running time (s).}
\begin{tabular}{c|cccccccc}
\hline
Algorithm            & \multicolumn{2}{c}{Trajectory PMBM} & \multicolumn{2}{c}{Trajectory MBM} & \multicolumn{2}{c}{Trajectory $\text{MBM}_{01}$} & $\delta$-GLMB (Gibbs) & LMB (Gibbs)    \\ \hline
Fixed-lag smoothing                    & w.o.                & w.                & w.o.                & w.               & w.o.                       & w.                      & w.o.             & w.o.      \\ \hline
GOSPA                & 150.02           & 150.02           & 148.98           & 148.98          & 149.33                  & 149.33                 & 151.94        & 155.21 \\
GOSPA-Localization   & 120.73           & 120.73           & 120.76           & 120.76          & 120.74                  & 120.74                 & 120.82        & 120.93 \\
GOSPA-Missed         & 68.10            & 68.10            & 66.24            & 66.24           & 67.54                   & 67.54                  & 63.71         & 57.72  \\
GOSPA-falsed         & 65.65            & 65.65            & 64.04            & 64.04           & 63.70                   & 63.70                  & 68.40         & 77.19  \\ \hline
LP trajectory metric & 141.91           & 128.25           & 141.02           & 127.15          & 141.04                  & 127.16                 & 167.50        & 168.85 \\
LP-Localization      & 123.23           & 101.72           & 123.40           & 101.87          & 123.35                  & 101.72                 & 123.01        & 123.01 \\
LP-Missed            & 98.10            & 98.10            & 93.81            & 93.81           & 93.89                   & 93.89                  & 131.80       & 128.21 \\
LP-False             & 56.38            & 56.38            & 62.56            & 62.56           & 63.19                   & 63.19                  & 107.46        & 114.76 \\
LP-Track switch      & 9.68             & 9.68             & 7.73             & 7.73            & 6.00                    & 6.00                   & 22.73         & 30.79  \\ \hline
Average running time (s) & 4.41             & 4.61             & 8.57             & 8.90            & 10.29                   & 10.50                  & 12.87         & 2.27   \\ \hline
\end{tabular}
\label{tab:scenario1}
\end{table*}

\begin{figure*}
    \centering
    \includegraphics[width=0.24\textwidth]{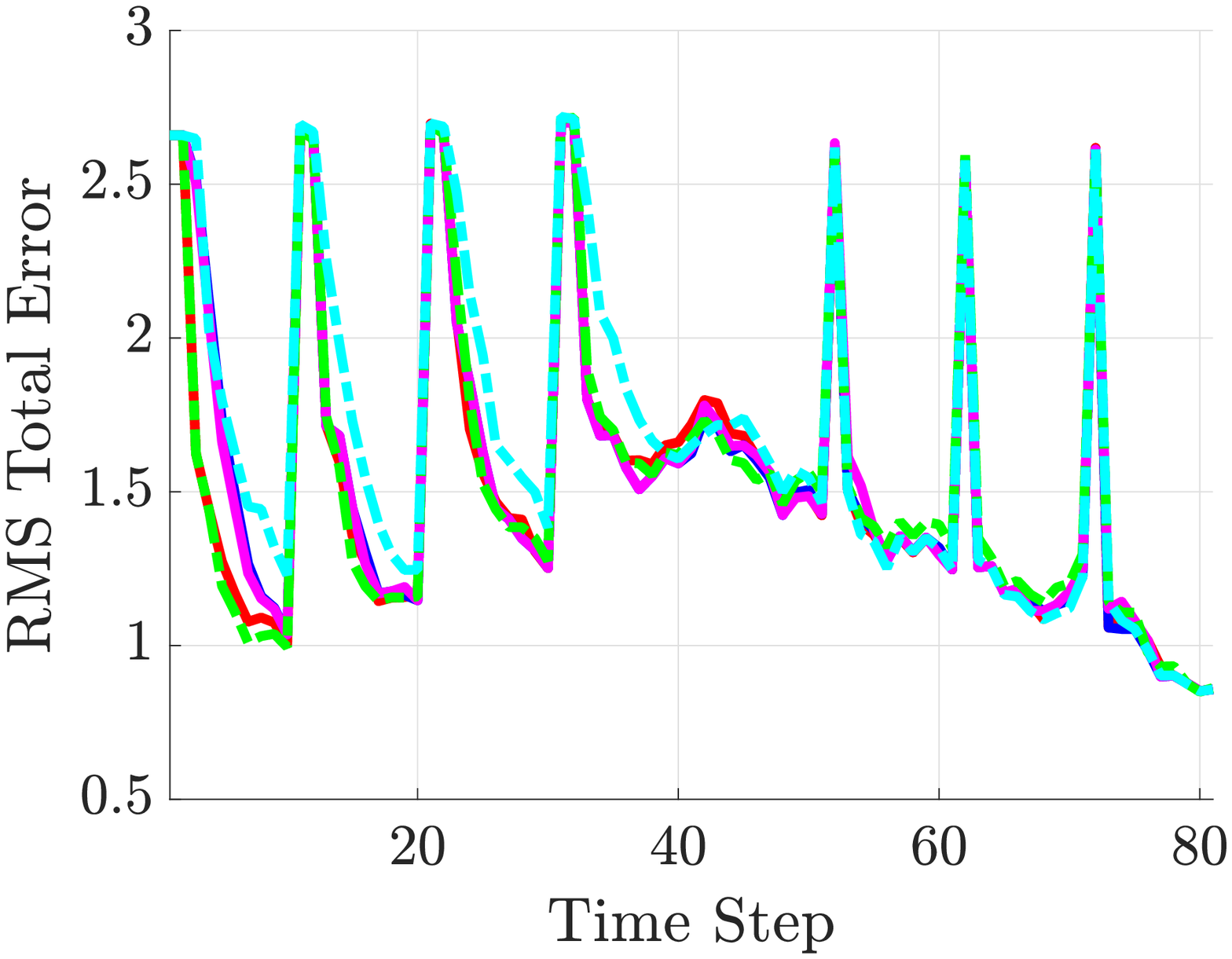}
    \includegraphics[width=0.24\textwidth]{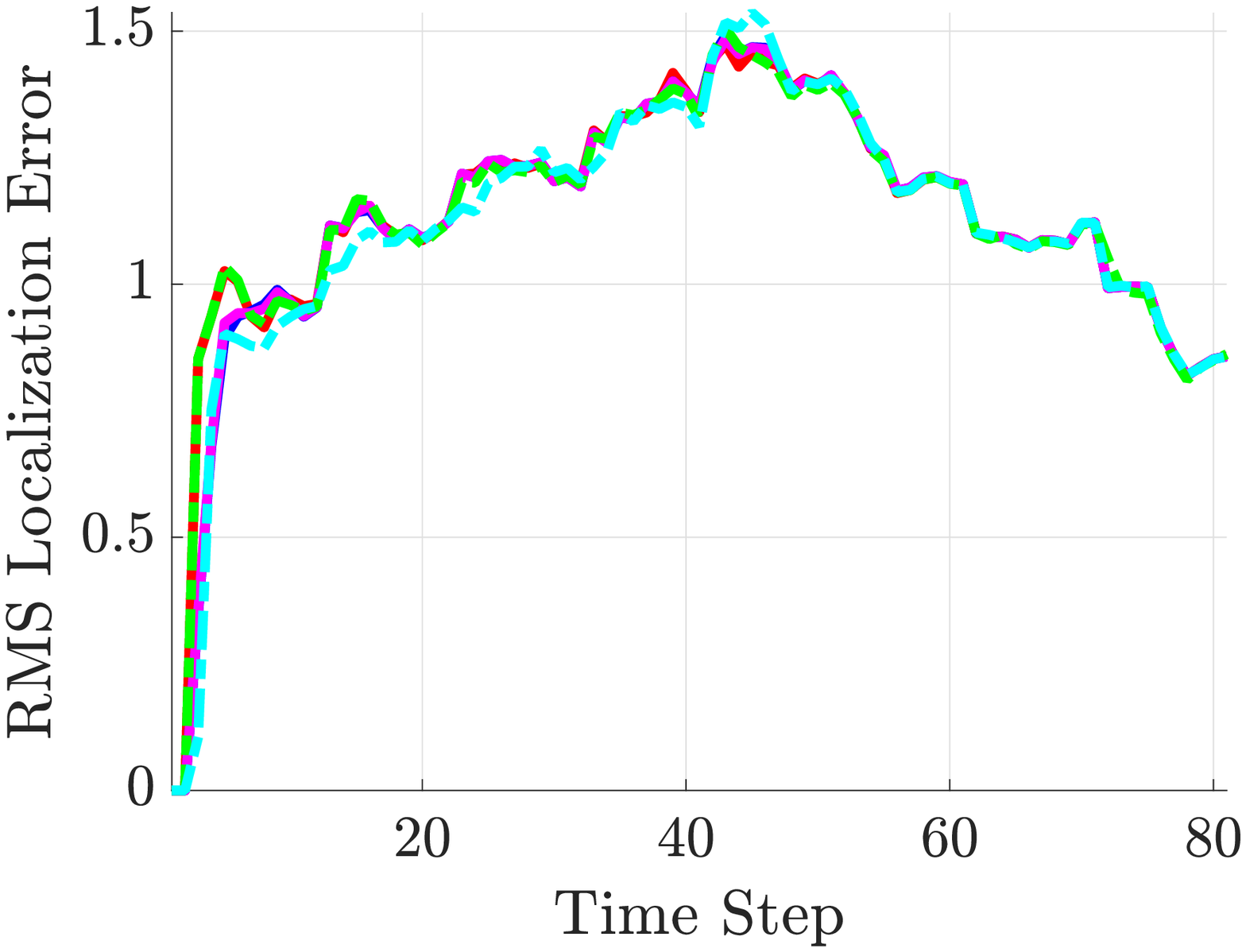}
    \includegraphics[width=0.24\textwidth]{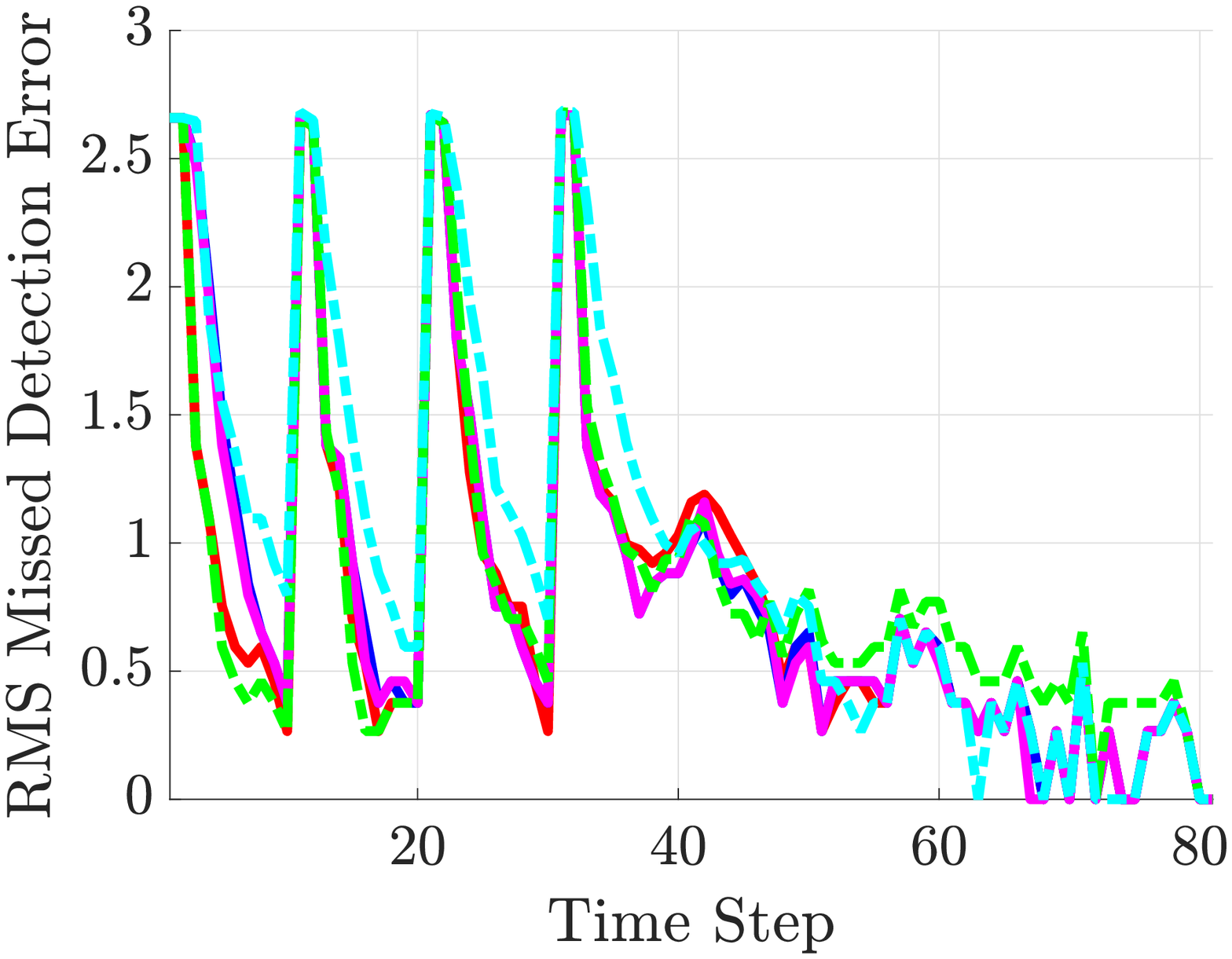}
    \includegraphics[width=0.24\textwidth]{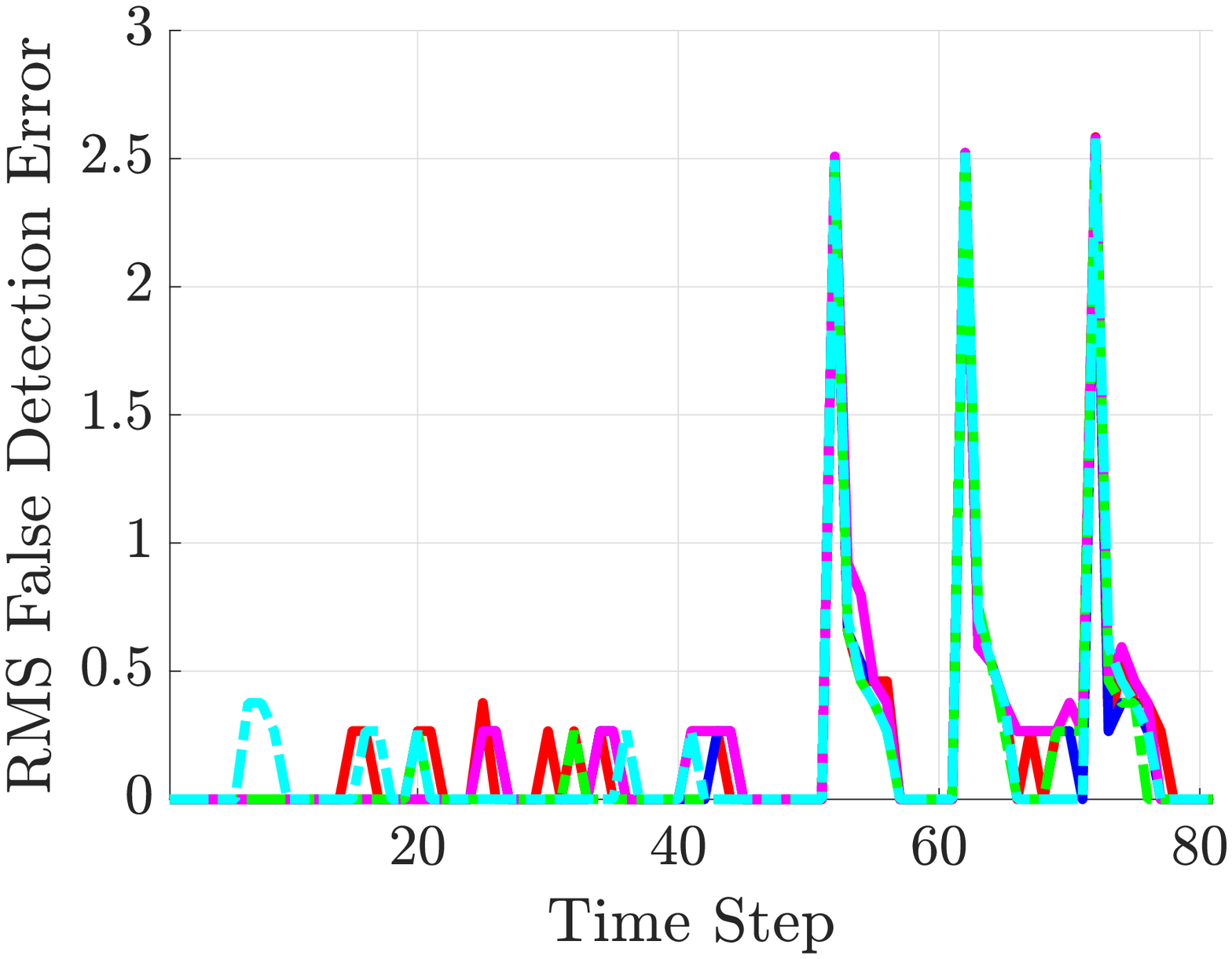}
    \caption{Average target state estimation error in Scenario 2 evaluated using the GOSPA metric. The lines show the RMS error averaged over 100 Monte Carlo runs. Legend: trajectory PMBM filter (w.o. smoothing) (\textcolor{red}{red}), trajectory MBM filter (w.o. smoothing) (\textcolor{blue}{blue}), trajectory $\text{MBM}_{01}$ filter (w.o. smoothing) (\textcolor{magenta}{magenta}), $\delta$-GLMB (Gibbs) filter (\textcolor{green}{green}), LMB (Gibbs) filter (\textcolor{cyan}{cyan}).}
    \label{fig:gospa}
\end{figure*}

\begin{figure*}[!t]
    \centering
    \includegraphics[width=0.19\textwidth]{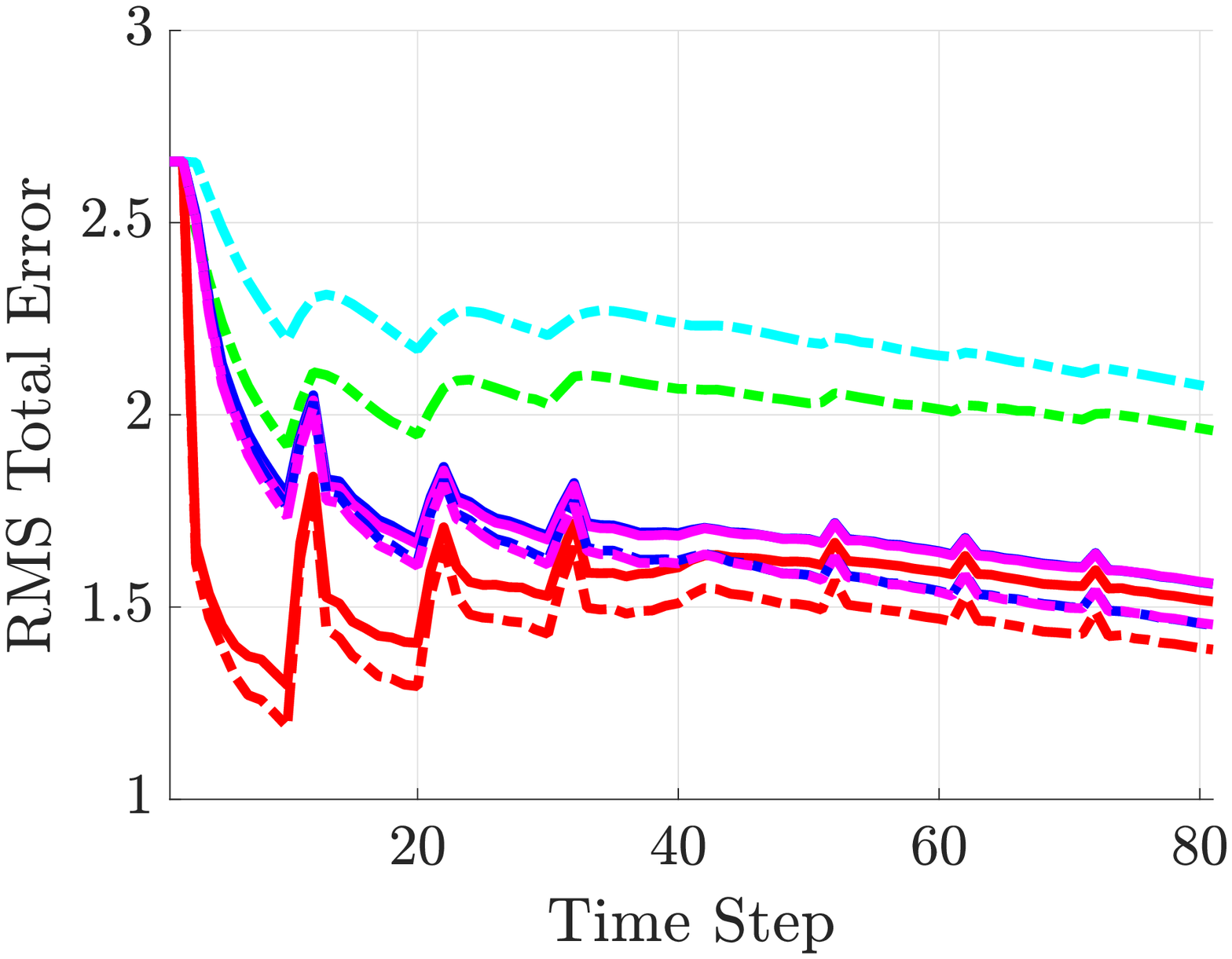}
    \includegraphics[width=0.19\textwidth]{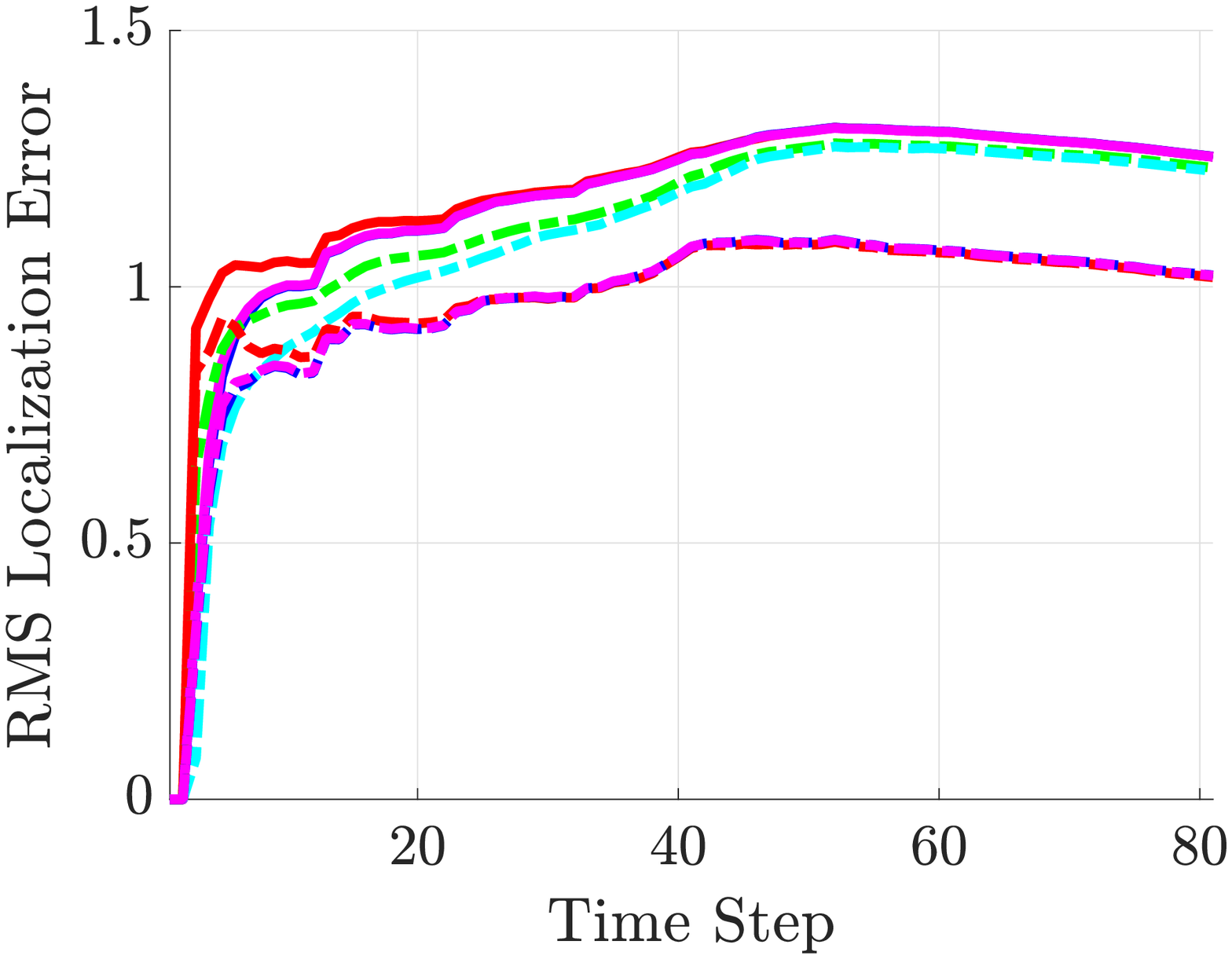}
    \includegraphics[width=0.19\textwidth]{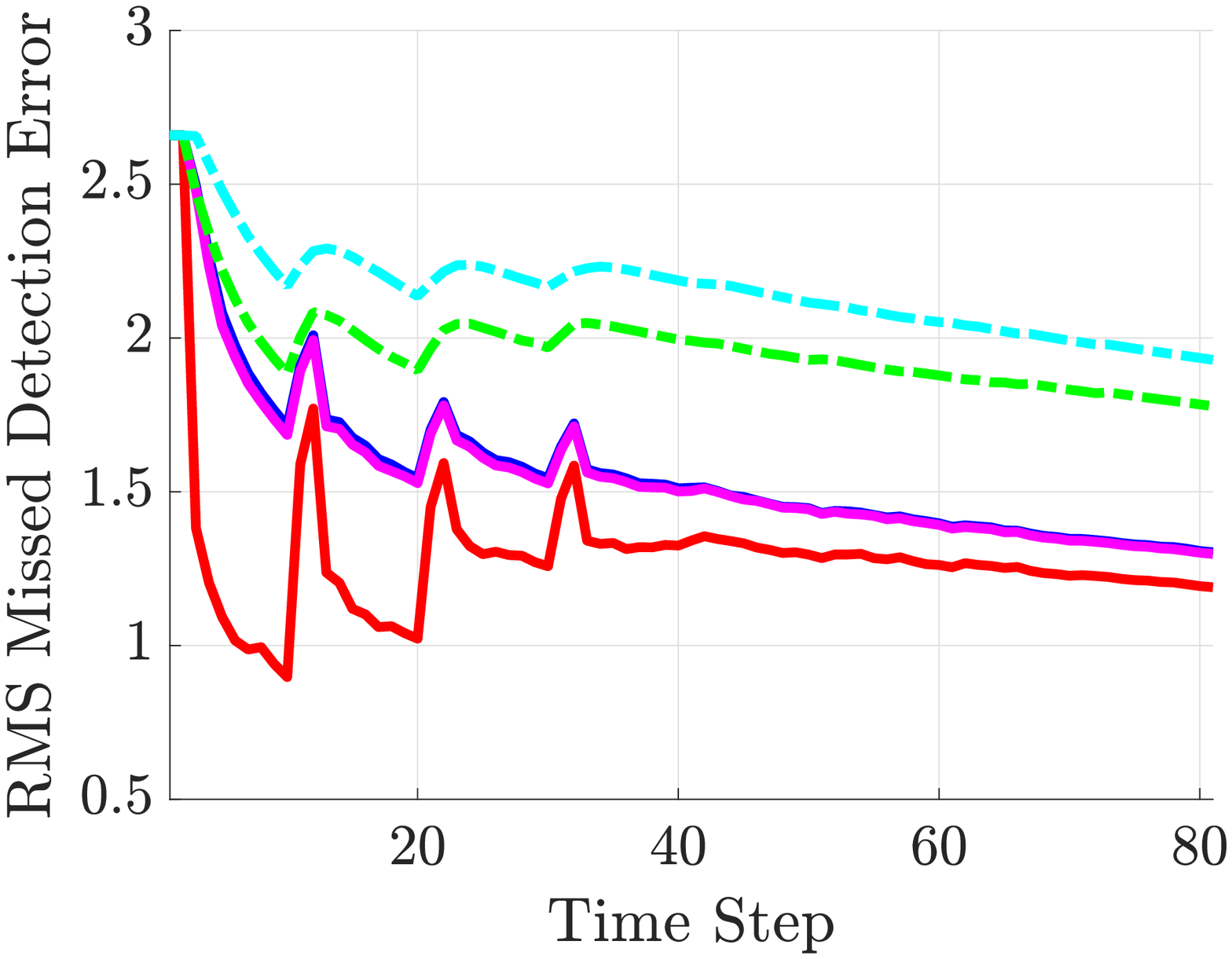}
    \includegraphics[width=0.19\textwidth]{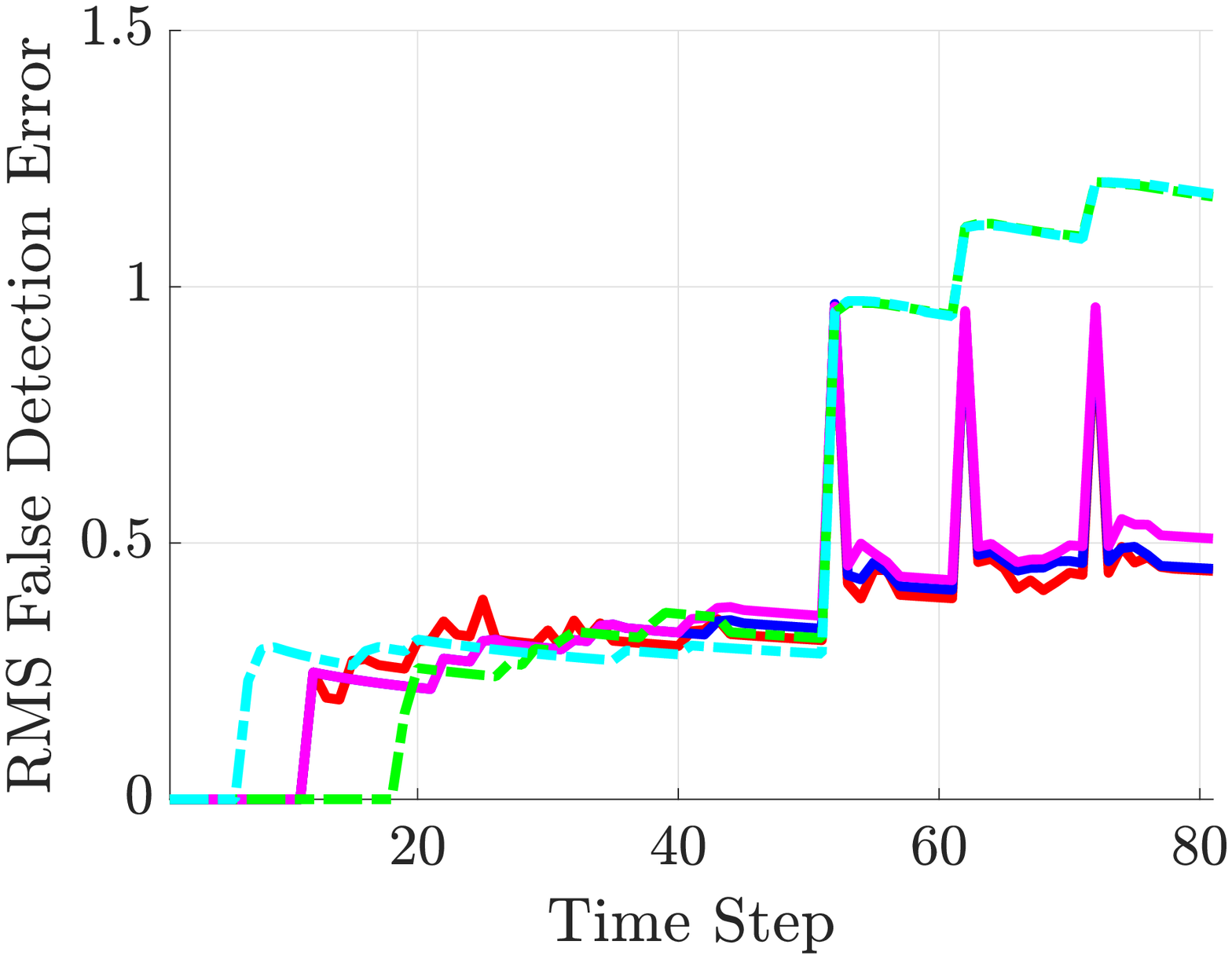}
    \includegraphics[width=0.19\textwidth]{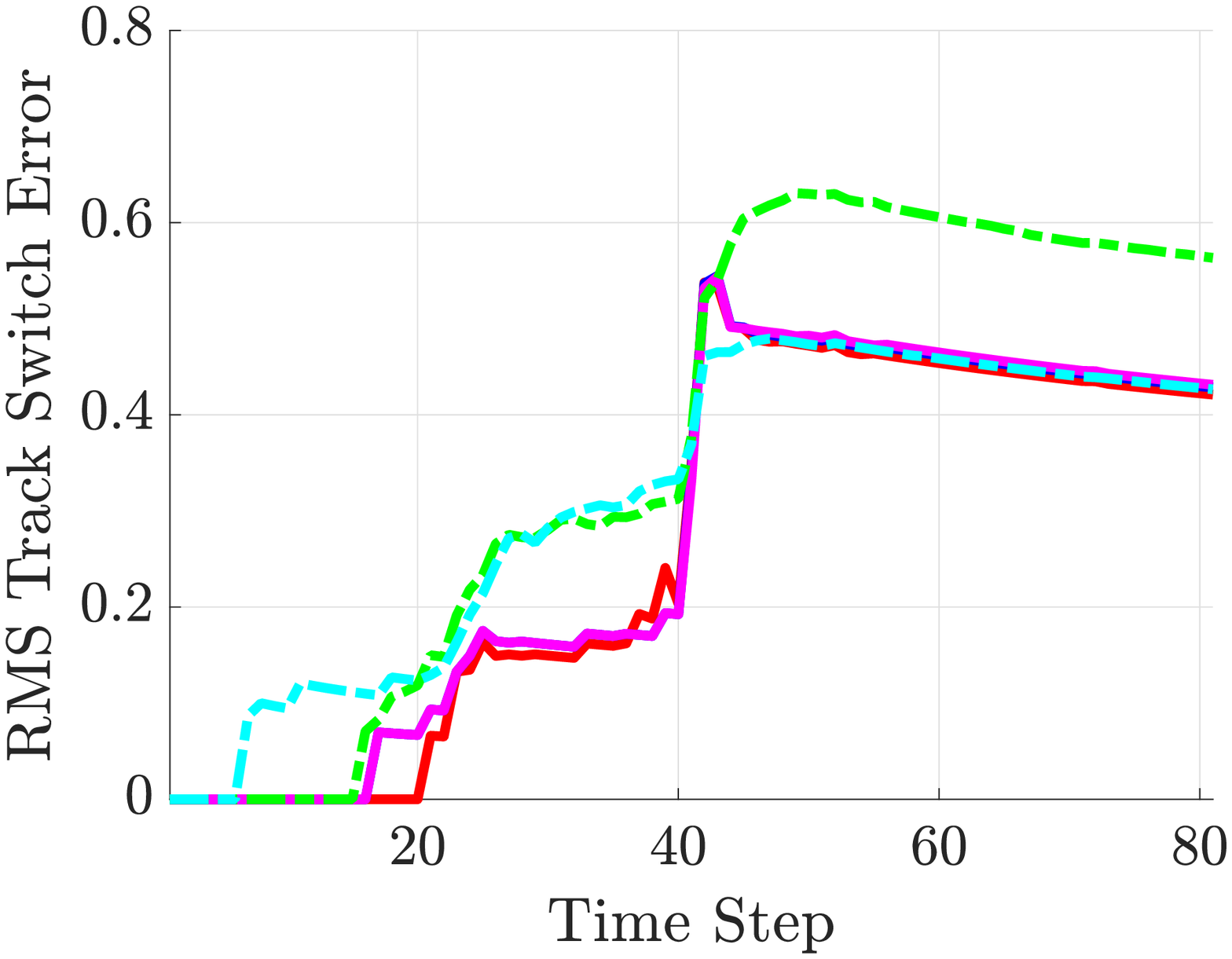}
    \caption{Average trajectory state estimation error in Scenario 2 evaluated using the trajectory metric \cite{trajectorymetric}. The lines show the RMS error averaged over 100 Monte Carlo runs. Legend: trajectory PMBM filter (w.o. smoothing) (\textcolor{red}{red solid line}), trajectory PMBM filter (w. smoothing) (\textcolor{red}{red dashdot line}), trajectory MBM filter (w.o. smoothing) (\textcolor{blue}{blue solid line}), trajectory MBM filter (w. smoothing) (\textcolor{blue}{blue dashdot line}), trajectory $\text{MBM}_{01}$ filter (w.o. smoothing) (\textcolor{magenta}{magenta solid line}), trajectory $\text{MBM}_{01}$ filter (w. smoothing) (\textcolor{magenta}{magenta dashdot line}), $\delta$-GLMB (Gibbs) filter (\textcolor{green}{green}), LMB (Gibbs) filter (\textcolor{cyan}{cyan}).}
    \label{fig:LPmetric}
\end{figure*}

\subsection{Results}
We perform 100 Monte Carlo runs and obtain the average root mean square (RMS) GOSPA error (order $p=2$, location error cut-off $c=10$, and $\alpha=2$), the average RMS trajectory estimation error (order $p=2$, location error cut-off $c=10$, switch cost $\gamma=2$), and the average running time, summed over 81 time steps. We apply the trajectory metric \cite{trajectorymetric} at each time step $k$, and normalise it by $\sqrt{k}$. This normalization allows a comparison of how the RMS metric evolves over time in the scenario, as opposed to only computing the metric at the final time step.

The comparison of different filters by the RMS GOSPA error and by the average running time\footnote{\texttt{MATLAB} implementations on a desktop with 3.0 GHz Intel Core i5 processor.} is shown in Table \ref{tab:scenario1} for Scenario 1, and in Figure \ref{fig:time_vs_error} for Scenario 2. We can see that the trajectory PMBM filter arguably has the best performance in terms of target state estimation error and computational complexity, especially in Scenario 2 with coalescence. By comparing the execution time of trajectory filters with and without fixed-lag smoothing (for the latest four target states), we can find that the running time of the implemented filters is dominated by their filtering recursions. 

For Scenario 1, the numerical values of the average RMS GOSPA and the trajectory estimation errors are given in Table \ref{tab:scenario1}. For Scenario 2, the average RMS GOSPA error and its decomposed values over time are illustrated in Figure \ref{fig:gospa}, and the average RMS trajectory estimation error and its decomposed values over time are illustrated in Figure \ref{fig:LPmetric}. Comparing the results of the two scenarios, we can find that when the birth process is less informative, i.e., a broad birth prior density, the trajectory PMBM filter exhibits lower estimation error than the trajectory MBM and $\text{MBM}_{01}$ filters.

While the differences in target state estimation error among different filters are not distinct in both scenarios, it is noticeable that trajectory filters yield much less trajectory estimation error than labelled RFS filters. The worse trajectory estimation performance of labelled RFS filters is a result of worse track continuity. There are two main drawbacks in forming trajectories by connecting target states with the same label: first, misdetections can lead to gaps in the trajectory formed by labelled estimates; second, physically unrealistic track switching, see \cite[Fig. 2]{continuityPMBM} for an example.

In addition, we can see that performing fixed-lag smoothing does not change the error due to missed/false detections and track switching; it mainly improves the localization error. This is expected since the choice of $N+L$ has a direct effect on the estimation of past states of the trajectories. From the results of the simulation study, we can conclude that the trajectory PMBM filter has the best tracking performance, and that the trajectory MBM filter is more efficient than the trajectory $\text{MBM}_{01}$ filter.

\section{\normalfont Conclusion}
In this paper, we have presented the trajectory MBM filter. We have also presented an efficient implementation of multi-scan trajectory PMBM, MBM and $\text{MBM}_{01}$ filters using $N$-scan pruning and dual decomposition. The performance of the presented multi-target trackers, applied with an efficient fixed-lag smoothing method, are evaluated in a simulation study. The simulation results show that the multi-scan trajectory PMBM filter has improved tracking performance over the trajectory MBM filter in terms of state/trajectory estimation error and computational time.

% This paper has proposed an efficient implementation of the PMBM trajectory filter. Compared with the existing PMBM target filter, the PMBM trajectory filter is able to estimate all target trajectories. Also, a simulation study shows that the PMBM trajectory filter has better performance than the $\delta$-GLMB filter in terms of state/trajectory estimation error and computational time in a scenario with coalescence.

% The next stage of this paper will involve combining the $L$-scan window approximation technique \cite{garcia2016trajectory} with the track-oriented $N$-scan pruning and studying how to perform merging and recycling in the PMBM trajectory filter.

% if have a single appendix:
%\appendix[Proof of the Zonklar Equations]
% or
%\appendix  % for no appendix heading
% do not use \section anymore after \appendix, only \section*
% is possibly needed

% use appendices with more than one appendix
% then use \section to start each appendix
% you must declare a \section before using any
% \subsection or using \label (\appendices by itself
% starts a section numbered zero.)
%

\appendices
\section{}

\label{ap:a}

In this appendix, we first review why FISST can be used for sets of trajectories. Then, we show how to define reference measures and measure theoretic integrals for sets of trajectories.

\subsection{Use of FISST for Sets of Trajectories}

In this subsection, we review why FISST can be used for sets of trajectories. The single trajectory space is locally compact, Hausdorff and second-countable (LCHS) \cite[App. A]{trackingbasedontrajectories}, where second-countable is also referred to as completely separable \cite{simmons1975topology}. LCHS spaces are often used in random set theory \cite{molchanov2005theory}, and LCHS is also the type of single-object space required by Mahler's FISST \cite[Sec. 2.2.2]{rfs}.

In particular, single object/measurement spaces that are the disjoint union of spaces of different dimensionalities, similarly to the single trajectory space, have previously been used in \text{Mahler's} FISST and RFS framework in \cite[Sec. 2.2.2]{rfs}, \cite[Sec. 11.6]{rfs} for variable state space cardinalized probability hypothesis density filters, and in \cite[Chap. 18]{rfs}, \cite{mahler2010cphd,mahler2011cphd} for RFS filters for unknown clutter. In addition, \cite[Sec. 3.5.3]{rfs} explicitly explains how the set integral is constructed for this type of space. Therefore, Mahler's FISST and RFS framework on its own enables us to perform inference on sets of trajectories. For completeness, we proceed to provide also the required measure theory to define probability densities.

\subsection{Measure Theoretic Integrals}

We begin by introducing some basic concepts in measure theory, for more details see, e.g., \cite{grimmett2001probability}, \cite[App. A]{vo2005sequential}. Consider a nonempty set $\mathcal{Y}$, the pair $(\mathcal{Y},\sigma(\mathcal{Y}))$, in which $\sigma(\mathcal{Y})$ denotes a $\sigma$-algebra of subsets of $\mathcal{Y}$, is called a measurable space. Given a topology space $\mathcal{Y}$, the Borel $\sigma$-algebra is the smallest $\sigma$-algebra of the subsets of $\mathcal{Y}$ containing the open sets of $\mathcal{Y}$ (or equivalently, by the closed sets of $\mathcal{Y}$). A set $\mathcal{B}$ is said to be measurable if $\mathcal{B}\in \sigma(\mathcal{Y})$. A function $f: \mathcal{Y}\rightarrow \mathbb{R}$ is said to be measurable if the inverse images of $\mathbb{R}$ under $f$ are measurable. The triple $(\mathcal{Y},\sigma(\mathcal{Y}),\mu)$ in which $\mu$ is a measure on $\sigma(\mathcal{Y})$ is called a measure space. 
% If $\mathcal{X}=\cup^{\infty}_{i=0}\mathcal{B}_i$ for some countable sequences of $\mathcal{B}_i\in\sigma(\mathcal{X})$ with $\mu(\mathcal{B}_i)<\infty$, then $\mu$ is said to be $\sigma$-finite.

The integral of a measurable function $f:\mathcal{Y}\rightarrow \mathbb{R}$, $\int f(y)\mu (dy)$, is defined as a limit of integrals of simple functions. The integral of $f$ over any measurable $\mathcal{B}\subset \mathcal{Y}$ is defined as 
\begin{equation}
    \int_{\mathcal{B}}f(y)\mu(dy) = \int \mathbf{1}_{\mathcal{B}}(y)f(y)\mu(dy),
\end{equation}
where $\mathbf{1}_{\mathcal{B}}$ denotes the indicator function $\mathbf{1}_{\mathcal{B}}(y)=1$ if $y\in\mathcal{B}$ and $\mathbf{1}_{\mathcal{B}}(y)=0$ otherwise.

\subsection{Measure Theoretic Integrals for Single Object LCHS Spaces}

In this subsection we explain how to define measure theoretic integrals for random finite sets whose single objects belong to LCHS spaces, following the steps in \cite[App. B]{vo2005sequential}. 

We denote an LCHS space as $E$. For instance, $E$ could denote the single object space $\mathcal{X}$ or the single trajectory space $\mathcal{T}_k$. We also let $\mathcal{F}(E)$ denote the collection of finite subsets of $E$\footnote{We would like to clarify that the topology on $\mathcal{F}(E)$ is the myopic of Math\'eron topology \cite{matheron1975random}, for which we require an LCHS space. To be precise, second-countability, not only separability as indicated in \cite[App. B]{vo2005sequential}, is required in the Math\'eron topology \cite[Sec. 1.1]{matheron1975random}, as it makes use of a countable base \cite[p. 1]{matheron1975random}.}. 

A common class of RFSs are the Poisson point processes. A Poisson point process $\Upsilon$ is an RFS that is characterized by the property that for any $k$ disjoint Borel subsets $S_1,...,S_k$ of $E$, the random variables $|\Upsilon\cap S_1|,...,|\Upsilon\cap S_k|$ are independent and have a Poisson distribution. The mean of the Poisson random variables $|\Upsilon\cap S_i|$ is denoted as $v_{\Upsilon}(S_i)$. The function $v_{\Upsilon}(\cdot)$ is a (unitless) measure on the Borel subsets of $E$ and is referred to as the intensity measure of $\Upsilon$. If the mapping from vectors to finite sets is denoted as $\chi:\uplus^{\infty}_{n=0}E^n\rightarrow\mathcal{F}(E)$, we have that $\chi((x_1,...,x_n)) = \{x_1,...,x_n\}$. Then, the probability distribution of $\Upsilon$ is \cite[App. B]{vo2005sequential}
\begin{equation}
    P_{\Upsilon}(\mathcal{B}) = e^{-v_{\Upsilon}(E)}\sum^{\infty}_{n=0}\frac{v^n_{\Upsilon}(\chi^{-1}(\mathcal{B})\cap E^n)}{n!},
\end{equation}
where $\mathcal{B}$ is a Borel subset of $\mathcal{F}(E)$, $\chi^{-1}$ is the inverse mapping of $\chi$, and $v^n_{\Upsilon}(\cdot)$ is the $n$-th product (unitless) Lebesgue measure of $v_{\Upsilon}(\cdot)$.

We define the measure $\mu(\cdot)$, on the Borel subsets of $\mathcal{F}(E)$, as 
\begin{equation}
    \mu(\mathcal{B}) = \sum^{\infty}_{n=0}\frac{v^n_{\Upsilon}(\chi^{-1}(\mathcal{B})\cap E^n)}{n!},
    \label{eq:measureBorel}
\end{equation}
which is proportional to the probability distribution $P_{\Upsilon}(\cdot)$. The integral of a measurable function $f:\mathcal{F}(E)\rightarrow \mathbb{R}$ w.r.t. the measure $\mu(\cdot)$ is then \cite[App. B]{vo2005sequential},
\begin{multline}
    \int_{\mathcal{B}}f(\mathbf{X})\mu(d\mathbf{X}) =\\ \sum_{n=0}^{\infty}\frac{1}{n!}\int_{\chi^{-1}(\mathcal{B})\cap E^n} f(\{x_1,...,x_n\})v^n_{\Upsilon}(dx_1\cdot\cdot\cdot dx_n).
    \label{eq:settrajectoryintegral}
\end{multline}

\subsection{Reference Measure for Sets of Trajectories}

In the previous subsection, we explained how to define a measure theoretic integral w.r.t. a measure $\mu(\cdot)$ on the Borel subsets of $\mathcal{F}(E)$ in terms of a measure $v_{\Upsilon}(\cdot)$ on the Borel subsets of $E$. We proceed to choose a specific measure $v_{\Upsilon}(\cdot)$ when $E$ is the single trajectory space $\mathcal{T}_k = \uplus_{(\beta,\varepsilon)\in I_k}\{\beta\}\times\{\varepsilon\}\times\mathcal{X}^{\varepsilon-\beta+1}$ and $\mathcal{X}=\mathbb{R}^n$. This will allow us to write the measure theoretic integrals for sets of trajectories in terms of standard Lebesgue integrals and establish the correspondence with Mahler's set integral (\ref{eq:setintegral}).

We first denote the units of the hyper-volume in the single target space $\mathcal{X}$ as $K$. For example, if the single target state is $[p_x,v_x]$ with $p_x$ being measured in meters $(m)$ and $v_x$ being measured in meters per second $(m/s)$, then, $K=m^2/s$.

Given a Borel subset $S$ of $\mathcal{T}_k$, which can be written as $S = \uplus_{(\beta,\varepsilon)\in I_k}\{\beta\}\times\{\varepsilon\}\times S_{\varepsilon-\beta+1},~S_{\varepsilon-\beta+1}\subset \mathcal{X}^{\varepsilon-\beta+1}$, we choose the measure $v_{\Upsilon}(\cdot)$ in the single trajectory space as 
\begin{equation}
    v_{\Upsilon}(S) = \sum_{(\beta,\varepsilon)\in I_k}\frac{\lambda_{K^{{\varepsilon-\beta+1}}}(S_{{\varepsilon-\beta+1}})}{K^{\varepsilon-\beta+1}},
    \label{eq:measure}
\end{equation}
where $\lambda_{K^{\varepsilon-\beta+1}}(\cdot)$ represents the Lebesgue measure of $S_{\varepsilon-\beta+1}$ (with units $K^{\varepsilon-\beta+1}$). Therefore, $\frac{\lambda_{K^{{\varepsilon-\beta+1}}}(\cdot)}{K^{\varepsilon-\beta+1}}$ represents the unitless Lebesgue measure on $\mathcal{X}^{\varepsilon-\beta+1}$. The normalization of each term in (\ref{eq:measure}) by $K^{\varepsilon-\beta+1}$ is needed so that we can perform the sum; otherwise, the sum would consider terms with different units, which is erroneous. It is straightforward to check that (\ref{eq:measure}) is a measure on the Borel subsets of $\mathcal{T}_k$. That is, $v_{\Upsilon}(\cdot)$ meets the following three properties that define measures \cite{billingsley2008probability}:
\begin{enumerate}
    \item For any $S$, $v_{\Upsilon}(S)\geq 0$.
    \item $v_{\Upsilon}(\emptyset)=0$.
    \item If $S^1,S^2,...$ is a disjoint sequence, then $v_{\Upsilon}(\sum_{j=1}^{\infty}S^j)=\sum^{\infty}_{j=1}v_{\Upsilon}(S^j)$.
\end{enumerate}
It is straightforward that the first two properties hold. For the third one, we have
\begin{equation}
\begin{split} 
v_{\Upsilon}\left(\sum^{\infty}_{j=1}S^j\right) &= \sum_{(\beta,\varepsilon)\in I_k}\frac{\lambda_{K^{\varepsilon-\beta+1}}(\sum^{\infty}_{j=1}S^j_{\varepsilon-\beta+1})}{K^{\varepsilon-\beta+1}}\\
&= \sum^{\infty}_{j=1}\sum_{(\beta,\varepsilon)\in I_k}\frac{\lambda_{K^{\varepsilon-\beta+1}}(S^j_{\varepsilon-\beta+1})}{K^{\varepsilon-\beta+1}}\\
&= \sum^{\infty}_{j=1}v_{\Upsilon}(S^j),
\end{split}
\end{equation}
where we have applied that $\lambda_{K^{\varepsilon-\beta+1}}(\cdot)$ is a measure.

We substitute (\ref{eq:measure}) into (\ref{eq:settrajectoryintegral}) and integrate over the whole space, which implies that $\mathcal{B}$ satisfies that $\chi^{-1}(\mathcal{B})\cap\mathcal{T}_k^n=\mathcal{T}_k^n$. We have that
\begin{equation}
    \begin{split}
        \int &f(\mathbf{X})\mu(d\mathbf{X}) \\&= \sum^{\infty}_{n=0}\frac{1}{n!}\int _{\mathcal{T}^n_k} f(\{X_1,...,X_n\})v_{\Upsilon}^n(dX_1...dX_n)\\
        &=\sum^{\infty}_{n=0}\frac{1}{n!}\int_{\mathcal{T}_k}\dots \int_{\mathcal{T}_k^n} f(\{X_1,...,X_n\})v_{\Upsilon}(dX_1)...v_{\Upsilon}(dX_n)\\
        &=\sum^{\infty}_{n=0}\frac{1}{n!}\sum_{(\beta_1,\varepsilon_1)\in I_k}\dots\sum_{(\beta_n,\varepsilon_n)\in I_k}\int_{\mathcal{X}^{\varepsilon_1-\beta_1+1}\times...\times\mathcal{X}^{\varepsilon_n-\beta_n+1}}\\
        &~~~~f(\{(\beta_1,\varepsilon_1,x_1^{1:\varepsilon_1-\beta_1+1}),...,(\beta_n,\varepsilon_n,x_n^{1:\varepsilon_n-\beta_n+1})\})\\
        &~~~~\frac{\lambda_{K^{\varepsilon_1-\beta_1+1}}(dx_1^{1:\varepsilon_1-\beta_1+1})}{K^{\varepsilon_1-\beta_1+1}}\dots \frac{\lambda_{K^{\varepsilon_n-\beta_n+1}}(dx_n^{1:\varepsilon_n-\beta_n+1})}{K^{\varepsilon_n-\beta_n+1}}.
    \end{split}
\end{equation}
If we further rewrite $\lambda_{K^{\varepsilon_i-\beta_i+1}}(dx_i^{1:\varepsilon_i-\beta_i+1})$ as $dx_i^{1:\varepsilon_i-\beta_i+1}$ and abbreviate $\int_{\mathcal{X}^{\varepsilon_1-\beta_1+1}\times...\times\mathcal{X}^{\varepsilon_n-\beta_n+1}}$ as $\int$, then we have that
\begin{multline}
    \int f(\mathbf{X})\mu(d\mathbf{X}) = \sum^{\infty}_{n=0}\frac{1}{n!}\sum_{(\beta_1,\varepsilon_1)\in I_k}\dots\sum_{(\beta_n,\varepsilon_n)\in I_k}\int\dots\int\\f(\{(\beta_1,\varepsilon_1,x_1^{1:\varepsilon_1-\beta_1+1}),...,(\beta_n,\varepsilon_n,x_n^{1:\varepsilon_n-\beta_n+1})\})\\
    \frac{dx_1^{1:\varepsilon_1-\beta_1+1}}{K^{\varepsilon_1-\beta_1+1}}\dots \frac{dx_n^{1:\varepsilon_n-\beta_n+1}}{K^{\varepsilon_n-\beta_n+1}}.
\end{multline}

Therefore, for the reference measure $\mu(\cdot)$ in (\ref{eq:measureBorel}) and $v_{\Upsilon}(\cdot)$ in (\ref{eq:measure}), the measure theoretic integral corresponds to Mahler's set integral over sets of trajectories (\ref{eq:setintegral}) but normalising by the units of the differential $dx_1^{1:\varepsilon_1-\beta_1+1},...,dx_n^{1:\varepsilon_n-\beta_n+1}$, which are $K^{\varepsilon_1-\beta_1+1},...,K^{\varepsilon_n-\beta_n+1}$. The relation between set integrals and measure theoretic integrals is similar in the single target case \cite{vo2005sequential}. Therefore, if probability densities on sets of trajectories are defined w.r.t. the reference measure $\mu(\cdot)$, with $v_{\Upsilon}(\cdot)$ given by (\ref{eq:measure}), Mahler's multi-trajectory densities are equivalent to measure theoretic densities, except for the normalizing units. Note that if the state space has no units, the measure theoretic integral and Mahler's set integral are alike.

\section{}
\label{ap:pgfl}

In this appendix, we proceed to explain how to use probability generating
functionals (PGFLs), functional derivatives and the fundamental theorem
of multi-object calculus for RFSs of trajectories. These results are
important as PGFLs are useful tools to derive filters. First, the
prediction and update steps can be performed in the PGFL domain. Second,
the fundamental theorem of multi-object calculus indicates how
to recover the corresponding multi-object density from a PGFL, which
requires functional derivatives. We explain PGFLs in Section \ref{subsec:Probability-generating-functiona} and functional derivatives in Section \ref{subsec:Functional-derivatives}. In Section \ref{subsec:Fundamental-theorem-of},
we provide and prove the fundamental theorem of multi-object calculus
for RFSs of trajectories. 

\subsection{Probability Generating Functionals\label{subsec:Probability-generating-functiona}}

PGFLs for sets in LCHS spaces, such as the trajectory space, are defined in \cite[Sec. 4.2.4, 4.2.5]{rfs}. Let $h: \mathcal{T}_k\mapsto [0,1]$ be a test function defined on the trajectory state space $\mathcal{T}_k = \uplus_{(\beta,\varepsilon)\in I_k}\{\beta\}\times\{\varepsilon\}\times\mathcal{X}^{\varepsilon-\beta+1}$.
Let $\mathbf{X}$ be an RFS of trajectories with multi-trajectory density
$f\left(\cdot\right)$, then, its PGFL is

\begin{align}
G_{\mathbf{X}}\left[h\right]=\mathrm{E}\left[h^{\mathbf{X}}\right] & =\int h^{\mathbf{X}}f\left(\mathbf{X}\right)\delta\mathbf{X},\label{eq:PGFL}
\end{align}
where
\begin{align*}
h^{\mathbf{X}} & =\begin{cases}
\prod_{X\in\mathbf{X}}h\left(X\right), & \mathbf{X}\neq\emptyset\\
1. & \mathbf{X}=\emptyset
\end{cases}
\end{align*}
Note that both $h(X)$ and the PGFL are unitless functions. i.e., functions whose
output has no units. 

\subsection{Functional Derivatives\label{subsec:Functional-derivatives}}

In this section, we explain (Volterra) functional derivatives for
RFS of trajectories using FISST tools. We consider a unitless functional
$F\left[h\right]$ defined on unitless real-valued functions $h\left(X\right)$
with $X\in \mathcal{T}_k$, e.g., a PGFL. Then, using
FISST, the functional derivative of $F\left[h\right]$ with respect
to a finite subset $\mathbf{Y}\in \mathcal{F}(\mathcal{T}_k)$ is defined
to be \cite[Sec. 11.4]{mahler2007statistical}
\begin{align}
\frac{\delta F}{\delta\mathbf{Y}}\left[h\right] & =\begin{cases}
F\left[h\right], & \mathbf{Y}=\emptyset\\
\lim_{\varepsilon\rightarrow0}\frac{F\left[h+\varepsilon\delta_{Y}\right]-F\left[h\right]}{\varepsilon}, & \mathbf{Y}=\left\{ Y\right\} \\
\frac{\delta^{n}F}{\delta Y_{1}\ldots\delta Y_{n}}\left[h\right], & \mathbf{Y}=\left\{ Y_{1},...,Y_{n}\right\} 
\end{cases}\label{eq:functional_derivative}
\end{align}
where the Dirac delta on the single trajectory space is 
\begin{align*}
\delta_{\left(\beta^{\prime},\varepsilon^{\prime},y_{\beta^{\prime}:\varepsilon^{\prime}}\right)}\left(\beta,\varepsilon,x_{\beta:\varepsilon}\right) & =\begin{cases}
\delta\left(x_{\beta:\varepsilon}-y_{\beta^{\prime}:\varepsilon^{\prime}}\right), & \beta=\beta^{\prime},\varepsilon=\varepsilon^{\prime}\\
0, & \beta\neq\beta^{\prime},\varepsilon\neq\varepsilon^{\prime}
\end{cases}
\end{align*}
and we use the notational convention 
\begin{align*}
\frac{\delta F}{\delta\left\{ Y\right\} }\left[h\right] & =\frac{\delta F}{\delta Y}\left[h\right].
\end{align*}
Also, note that the Dirac delta on the single trajectory space
meets the following identity
\begin{align*}
\int\delta_{Y}\left(X\right)f\left(X\right)dX & =f\left(Y\right).
\end{align*}

We remark that the use of $\delta_{Y}$ as the input
of the functional is a tool of FISST that is not completely rigorous
\cite[p. 66]{rfs}, but admitted from a practical point
of view. Set derivatives can be defined in terms of functional derivatives
\cite[p. 67]{rfs}.

\subsection{Fundamental Theorem of Multi-Object Calculus\label{subsec:Fundamental-theorem-of}}

The fundamental theorem of multi-object calculus enables the recovery
of a multi-object density from its PGFL \cite[Sec. 3.5.1]{rfs}.
This result also applies to RFS of trajectories, and we provide a
proof for completeness.
\begin{theorem}
Given the PGFL $G_{\mathbf{X}}\left[h\right]$ of an RFS $\mathbf{X}$
of trajectories, we can recover its multi-trajectory density $f\left(\cdot\right)$
evaluated at $\mathbf{Y}$ as
\begin{align}
f\left(\mathbf{Y}\right) & =\left[\frac{\delta G_{\mathbf{X}}}{\delta\mathbf{Y}}\left[h\right]\right]_{h=0}.\label{eq:fundamental_theorem}
\end{align}
\end{theorem}
The proof of this theorem is direct for $\mathbf{Y}=\emptyset$ by
substituting (\ref{eq:functional_derivative}) into (\ref{eq:PGFL}).
For $\mathbf{Y}\neq\emptyset$, the theorem is a direct consequence
of the following lemma.
\begin{lemma}
\label{lem:Functional-derivative}The functional derivative of the
PGFL $G_{\mathbf{X}}\left[h\right]$ of an RFS $\mathbf{X}$ of trajectories
with respect to $\mathbf{Y}=\left\{ Y_{1},...,Y_{n}\right\} $ is
\begin{align}
\frac{\delta^{n}G_{\mathbf{X}}}{\delta Y_{1}\ldots\delta Y_{n}}\left[h\right] & =\int h^{\mathbf{X}}f\left(\left\{ Y_{1},...,Y_{n}\right\} \cup\mathbf{X}\right)\delta\mathbf{X},\label{eq:consecutive_derivatives}
\end{align}
where $f\left(\cdot\right)$ is its multi-trajectory density.
\end{lemma}
The proof of Lemma \ref{lem:Functional-derivative} is given in the
subsection \ref{subsec:Proof-consecutive_derivatives}. Then by substituting
$h=0$, we directly obtain (\ref{eq:fundamental_theorem}) for $\mathbf{Y}\neq\emptyset$.
We also have 
\begin{align*}
\left[\frac{\delta G_{\mathbf{X}}}{\delta Y}\left[h\right]\right]_{h=1} & =\int f\left(\left\{ Y\right\} \cup\mathbf{X}\right)\delta\mathbf{X},
\end{align*}
which represents the first-order moment, also called intensity and
probability hypothesis density, as required.

\subsubsection{Proof of Lemma \ref{lem:Functional-derivative}\label{subsec:Proof-consecutive_derivatives}}

In this section, we prove (\ref{eq:consecutive_derivatives}) by using
induction. In Part I of the proof, we prove (\ref{eq:consecutive_derivatives})
for $\mathbf{Y}=\left\{ Y\right\} $. Then, in Part II, we prove the
general case $\mathbf{Y}=\left\{ Y_{1},...,Y_{n}\right\} $.

\paragraph{Part I of the Proof}

For $\mathbf{Y}=\left\{ Y\right\} $, we proceed to prove that 
\begin{align*}
\frac{\delta G_{\mathbf{X}}}{\delta Y}\left[h\right] & =\int h^{\mathbf{X}}f\left(\left\{ Y\right\} \cup\mathbf{X}\right)\delta\mathbf{X}.
\end{align*}
For $\mathbf{Y}=\left\{ Y\right\} $, we have 
\begin{align*}
 & \frac{\delta G_{\mathbf{X}}}{\delta Y}\left[h\right]\\
 & =\lim_{\epsilon\rightarrow0}\frac{G_{\mathbf{X}}\left[h+\epsilon\delta_{Y}\right]-G_{\mathbf{X}}\left[h\right]}{\epsilon}\\
 & =\lim_{\epsilon\rightarrow0}\frac{\int\left[h+\epsilon\delta_{Y}\right]^{\mathbf{X}}f\left(\mathbf{X}\right)\delta\mathbf{X}-\int\left[h\right]^{\mathbf{X}}f\left(\mathbf{X}\right)\delta\mathbf{X}}{\epsilon}\\
 & =\lim_{\epsilon\rightarrow0}\frac{\sum_{n=1}^{\infty}\frac{1}{n!}\int f\left(\left\{ X_{1},...,X_{n}\right\} \right)\times\cdot\cdot\cdot}{\epsilon}\\& \frac{\times\left[\prod_{j=1}^{n}\left[h\left(X_{j}\right)+\epsilon\delta_{Y}\left(X_{j}\right)\right]-\prod_{j=1}^{n}h\left(X_{j}\right)\right]dX_{1:n}}{\epsilon},
\end{align*}
where $X_{1:n}=\left(X_{1},...,X_{n}\right)$. The limit can be computed by applying L'H\^{o}pital's rule and taking derivatives with respect to
$\epsilon$. This results in
\begin{align*}
 & \frac{\delta G_{\mathbf{X}}}{\delta Y}\left[h\right]\\
 & =\lim_{\epsilon\rightarrow0}\sum_{n=1}^{\infty}\frac{1}{n!}\int\sum_{j=1}^{n}\left[\delta_{Y}\left(X_{j}\right)\prod_{i=1:i\neq j}^{n}h\left(X_{i}+\epsilon\delta_{Y}\left(X_{i}\right)\right)\right]\\
 & \quad\times f\left(\left\{ X_{1},...,X_{n}\right\} \right)dX_{1:n}\\
 & =\sum_{n=1}^{\infty}\frac{1}{n!}\sum_{j=1}^{n}\int\left[\delta_{Y}\left(X_{j}\right)\prod_{i=1:i\neq j}^{n}h\left(X_{i}\right)\right]\\
 & \quad\times f\left(\left\{ X_{1},...,X_{n}\right\} \right)dX_{1:n}.
\end{align*}
The inner integral is the same for every $j$, so we can write
\begin{align*}
 & \frac{\delta G_{\mathbf{X}}}{\delta\left\{ Y\right\} }\left[h\right]\\
 & =\sum_{n=1}^{\infty}\frac{1}{n!}n\int\left[\delta_{Y}\left(X_{1}\right)\prod_{i=2}^{n}h\left(X_{i}\right)\right]\\
 & \quad\times f\left(\left\{ X_{1},...,X_{n}\right\} \right)dX_{1:n}\\
 & =\sum_{n=1}^{\infty}\frac{1}{\left(n-1\right)!}\int\left[\prod_{i=2}^{n}h\left(X_{i}\right)\right]f\left(\left\{ Y,X_{2},...,X_{n}\right\} \right)dX_{2:n}.
\end{align*}
We further make the change of variables $m=n-1$ and $X_{1:m}^{*}=X_{2:n}$
in the previous equation, which yields
\begin{align}
 & \frac{\delta G_{\mathbf{X}}}{\delta\left\{ Y\right\} }\left[h\right]\nonumber \\
 & =\sum_{m=0}^{\infty}\frac{1}{m!}\int\left[\prod_{i=1}^{m}h\left(X_{i}^{*}\right)\right]f\left(\left\{ Y\right\} \cup\left\{ X_{1}^{*},...,X_{m}^{*}\right\} \right)dX_{1:m}^{*}\nonumber \\
 & =\int h^{\mathbf{X}}f\left(\left\{ Y\right\} \cup\mathbf{X}\right)\delta\mathbf{X}.\label{eq:PGFL_proof_one_element}
\end{align}

\paragraph{Part II of the Proof}

We proceed to prove (\ref{eq:consecutive_derivatives}) by induction.
We assume that 
\begin{align}
\frac{\delta^{n-1}G_{\mathbf{X}}}{\delta Y_{1}\ldots\delta Y_{n-1}}\left[h\right] & =\int h^{\mathbf{X}}f\left(\left\{ Y_{1},...,Y_{n-1}\right\} \cup\mathbf{X}\right)\delta\mathbf{X}
\end{align}
holds and the proceed to prove (\ref{eq:consecutive_derivatives}).
Note that the relation holds for $n=1$, as proved in the previous
section. We denote 
\begin{align*}
L\left[h\right] & =\int h^{\mathbf{X}}l\left(\mathbf{X}\right)\delta\mathbf{X},
\end{align*}
where
\begin{align*}
l\left(\mathbf{X}\right) & =f\left(\left\{ Y_{1},...,Y_{n-1}\right\} \cup\mathbf{X}\right).
\end{align*} Then, by making use of (\ref{eq:PGFL_proof_one_element}), we obtain
\begin{align*}
\frac{\delta^{n}G_{\mathbf{X}}}{\delta Y_{1}\ldots\delta Y_{n}}\left[h\right] & =\frac{\delta}{\delta Y_{n}}L\left[h\right]\\
 & =\int h^{\mathbf{X}}l\left(\left\{ Y_{n}\right\} \cup\mathbf{X}\right)\delta\mathbf{X}\\
 & =\int h^{\mathbf{X}}f\left(\left\{ Y_{1},...,Y_{n}\right\} \cup\mathbf{X}\right)\delta\mathbf{X}.
\end{align*}
This result completes the proof of Lemma \ref{lem:Functional-derivative}.
 
\section{}

\label{ap:b}

In this appendix, we present the $\text{MBM}_{01}$ filtering recursions for both the set of current trajectories and the set of all trajectories. The $\text{MBM}_{01}$ filtering recursions for the set of all trajectories was first given in \cite{trackingbasedontrajectories}; they are presented here for completeness.

\subsection{Prediction Step for the Set of Current Trajectories}
The prediction step is given in the theorem below.
\begin{theorem}
    Assume that the distribution from the previous time step $f_{k-1|k-1}(\mathbf{X}_{k-1})$ is given by (\ref{eq:mbm}) with $r^{i,a^i}_{k-1|k-1}\in\{0,1\}$, that the transition model is (\ref{eq:transition_current}), and that the birth model is a trajectory multi-Bernoulli RFS with $n^b_k$ Bernoulli components, each of which has density given by (\ref{eq:mbirth}). Then the predicted distribution for the next step $f_{k|k-1}(\mathbf{X}_{k})$ is given by (\ref{eq:mbm}) with $r^{i,a^i}_{k|k-1}\in\{0,1\}$ and $n_{k|k-1}=n_{k-1|k-1}+n^b_k$. For tracks continuing from previous time ($i\in\{1,...,n_{k-1|k-1}\}$), a hypothesis is included for each combination of a hypothesis from a previous time and either a survival or a death. For new tracks ($i\in\{n_{k-1|k-1}+l\}$, $l\in\{1,...,n^b_k\}$), a hypothesis is included for each combination of a Bernoulli component in the multi-Bernoulli birth density and either born or not born. The number of hypotheses therefore becomes $h^i_{k|k}=2(h^i_{k|k-1}+n^b_k)$.\footnote{A hypothesis at the previous time with $r^{i,a^i}_{k-1|k-1}=0$ would be removed by setting its hypothesis weight to zero. For simplicity, the hypothesis numbering does not account for this exclusion.} For survival hypotheses ($i\in\{1,...,n_{k-1|k-1}\}$, $a^i\in\{1,...,h_{k-1|k-1}\}$), if $r^{i,a^i}_{k-1|k-1}=1$, the parameters are
    \begin{subequations}
        \begin{align}            w^{i,a^i}_{k|k-1}&=w^{i,a^i}_{k-1|k-1}\langle f^{i,a^i}_{k-1|k-1};P^S_{k-1}\rangle,\\
            r^{i,a^i}_{k|k-1} &= 1,\\
            f^{i,a^i}_{k|k-1}(X) &= \langle f^{i,a^i}_{k-1|k-1};\pi^c\rangle.
        \end{align}
    \end{subequations}
    If $r^{i,a^i}_{k-1|k-1}=0$, the parameters are
    \begin{subequations}
        \begin{align}            
        r^{i,a^i}_{k|k-1} &= 0,\\
        w^{i,a^i}_{k|k-1} &= 0.
        \end{align}
    \end{subequations}
    For death hypotheses ($i\in\{1,...,n_{k-1|k-1}\}$, $a^i = \tilde{a}^i+h^i_{k-1|k-1}$, $\tilde{a}^i\in\{1,...,h^i_{k-1|k-1}\}$), the parameters are
    \begin{subequations}
        \begin{align}
         w^{i,a^i}_{k|k-1}&=w^{i,a^i}_{k-1|k-1}\langle f^{i,a^i}_{k-1|k-1};1-P^S_{k-1}\rangle,\\
            r^{i,a^i}_{k|k-1} &= 0.
        \end{align}
    \end{subequations}
    For birth hypotheses ($i\in\{n_{k-1|k-1}+l\}$, $l\in\{1,...,n^b_k\}$), the parameters are:
    \begin{subequations}
        \begin{align}
            \mathcal{M}^{k-1}(i,1) &= \emptyset,\\
            w^{i,1}_{k|k-1}&=r^{b,l}_{k},\\
            r^{i,1}_{k|k-1} &= 1,\\
            f^{i,1}_{k|k-1}(X) &= f^{B,l}_k(X).
        \end{align}
        \label{eq:mbm01born}
    \end{subequations}
    \\*
    For non-birth hypotheses ($i\in\{n_{k-1|k-1}+l\}$, $l\in\{1,...,n^b_k\}$), the parameters are:
    \begin{subequations}
        \begin{align}
            \mathcal{M}^{k-1}(i,2) &= \emptyset,\\
            w^{i,2}_{k|k-1}&=1-r^{b,l}_{k},\\
            r^{i,2}_{k|k-1} &= 0.
        \end{align}
        \label{eq:mbm01notborn}
    \end{subequations}
    \label{theorem4}
\end{theorem}

Compared to the corresponding prediction step (\ref{eq:mbmpredictexist}), (\ref{eq:newberparas}) in the trajectory MBM filter, the $\text{MBM}_{01}$ parameterization entails an exponential increase in the number of global hypotheses. 

\subsection{Prediction Step for the Set of All Trajectories}
The prediction step is given in the theorem below.
\begin{theorem}
    Assume that the distribution from the previous time step $f_{k-1|k-1}(\mathbf{X}_{k-1})$ is given by (\ref{eq:mbm}) with $r^{i,a^i}_{k-1|k-1}\in\{0,1\}$, that the transition model is (\ref{eq:transition_all}), and that the birth model is a trajectory multi-Bernoulli RFS with $n^b_k$ Bernoulli components, each of which has density given by (\ref{eq:mbirth}). Then the predicted distribution for the next step $f_{k|k-1}(\mathbf{X}_{k})$ is given by (\ref{eq:mbm}), with $r^{i,a^i}_{k|k-1}\in\{0,1\}$ and $n_{k|k-1}=n_{k-1|k-1}+n^b_k$. For tracks continuing from previous time ($i\in\{1,...,n_{k-1|k-1}\}$), the number of hypotheses remains the same. For new tracks ($i\in\{n_{k-1|k-1}+l\}$, $l\in\{1,...,n^b_k\}$), a hypothesis is included for each combination of a Bernoulli component in the multi-Bernoulli birth density and either born or not born. The number of hypotheses therefore becomes $h^i_{k|k}=h^i_{k|k-1}+2n^b_k$.\\
    For hypotheses in tracks continuing from previous time ($i\in\{1,...,n_{k-1|k-1}\}$, $a^i\in\{1,...,h_{k-1|k-1}\}$), the parameters are
    \begin{subequations}
        \begin{align}
         w^{i,a^i}_{k|k-1}&=w^{i,a^i}_{k-1|k-1}~\forall~a^i,\\
            r^{i,a^i}_{k|k-1} &= 1,\\
            f^{i,a^i}_{k|k-1}(X) &= \langle f^{i,a^i}_{k-1|k-1};\pi^a \rangle~\forall~a^i.
            \label{eq:mbm01_alltra_prediction}
        \end{align}
    \end{subequations}
    For new tracks ($i\in\{n_{k-1|k-1}+l\}$, $l\in\{1,...,n^b_k\}$), the parameters of $\text{MBM}_{01}$ parameterization are the same as (\ref{eq:mbm01born}) and (\ref{eq:mbm01notborn}).
    \label{theorem5}
\end{theorem}

\subsection{Update Step}
The update step is given in the theorem below.
\begin{theorem}
    Assume that the predicted distribution $f_{k|k-1}(\mathbf{X}_{k})$ is given by (\ref{eq:mbm}) with $r^{i,a^i}_{k|k-1}\in\{0,1\}$, that the measurement model is (\ref{eq:measurementmodel}), and that the measurement set at time step $k$ is $\mathbf{z}_k=\{z^1_k,...,z^{m_k}_k\}$. Then the updated distribution $f_{k|k}(\mathbf{X}_{k})$ is given by (\ref{eq:mbm}), with $r^{i,a^i}_{k|k}\in\{0,1\}$ and $n_{k|k} = n_{k|k-1}$. For each track ($i\in\{1,...,n_{k|k}\}$), a hypothesis is included for each combination of a hypothesis from a previous time with $r^{i,a^i}_{k|k-1}=1$ and either a misdetection or an update using one of the $m_k$ new measurements, such that the number of hypotheses becomes $h^i_{k|k}=h^i_{k|k-1}(1+m_k)$. \footnote{A hypothesis at the previous time with $r^{i,a^i}_{k|k-1}=0$ must not be updated. For simplicity, the hypothesis numbering does not account for this exclusion.} For misdetection hypotheses ($i\in\{1,...,n_{k|k}\}, a^i\in\{1,...,h_{k|k-1}\}$) with $r^{i,a^i}_{k|k-1}=1$, the parameters are
    \begin{subequations}
        \begin{align}
            \mathcal{M}^k(i,a^i) &= \mathcal{M}^{k-1}(i,a^i),\\
            w^{i,a^i}_{k|k} &=w^{i,a^i}_{k|k-1}\left(1-\left\langle  f^{i,a^i}_{k|k-1};P^D\right\rangle \right),\\
            r^{i,a^i}_{k|k} &=1,\\
            f^{i,a^i}_{k|k}(X) &= \frac{ (1-P^D_k(X))f^{i,a^i}_{k|k-1}(X) }{\left\langle f^{i,a^i}_{k|k-1};1-P^D \right\rangle}.
        \end{align}
    \end{subequations}
    For hypotheses updating tracks ($i\in\{1,...,n_{k|k}\}$, $a^i = \tilde{a}^i+h^i_{k|k-1}j$, $\tilde{a}^i\in\{1,...,h^i_{k|k-1}\}$, $j\in\{1,...,m_k\}$, i.e., the previous hypothesis $\tilde{a}^i$, updated with measurement $z^j_k$) with $r^{i,a^i}_{k|k-1}=1$, the parameters are
    \begin{subequations}
        \begin{align}
            \mathcal{M}^k(i,a^i) &= \mathcal{M}^{k-1}(i,\tilde{a}^i)\cup\{(k,j)\},\\
            w^{i,a^i}_{k|k} &=\frac{w^{i,a^i}_{k|k-1}\left\langle f^{i,\tilde{a}^i}_{k|k-1};\varphi(z^j_k|\cdot)P^D \right\rangle}{\lambda^{\text{FA}}(z^j_k)},\\
            r^{i,a^i}_{k|k} &= 1,\\
            f^{i,a^i}_{k|k}(X) &= \frac{\varphi(z^j_k|X)P^D_k(X)f^{i,\tilde{a}^i}_{k|k-1}(X)}{\left\langle f^{i,\tilde{a}^i}_{k|k-1};\varphi(z^j_k|\cdot)P^D_k\right\rangle}.
        \end{align}
    \end{subequations}
    \label{theorem6}
\end{theorem}

% Appendix one text goes here.

% you can choose not to have a title for an appendix
% if you want by leaving the argument blank
% \section{}
% Appendix two text goes here.

% use section* for acknowledgment
% \section*{Acknowledgment}

% The authors would like to thank...

% Can use something like this to put references on a page
% by themselves when using endfloat and the captionsoff option.
\ifCLASSOPTIONcaptionsoff
  \newpage
\fi

% trigger a \newpage just before the given reference
% number - used to balance the columns on the last page
% adjust value as needed - may need to be readjusted if
% the document is modified later
%\IEEEtriggeratref{8}
% The "triggered" command can be changed if desired:
%\IEEEtriggercmd{\enlargethispage{-5in}}

% references section

% can use a bibliography generated by BibTeX as a .bbl file
% BibTeX documentation can be easily obtained at:
% http://mirror.ctan.org/biblio/bibtex/contrib/doc/
% The IEEEtran BibTeX style support page is at:
% http://www.michaelshell.org/tex/ieeetran/bibtex/
%\bibliographystyle{IEEEtran}
% argument is your BibTeX string definitions and bibliography database(s)
%\bibliography{IEEEabrv,../bib/paper}
%
% <OR> manually copy in the resultant .bbl file
% set second argument of \begin to the number of references
% (used to reserve space for the reference number labels box)
% \begin{thebibliography}{1}

% \bibitem{IEEEhowto:kopka}
% H.~Kopka and P.~W. Daly, \emph{A Guide to \LaTeX}, 3rd~ed.\hskip 1em plus
%   0.5em minus 0.4em\relax Harlow, England: Addison-Wesley, 1999.

% \end{thebibliography}
\bibliographystyle{IEEEtran}
\bibliography{main}

% biography section
% 
% If you have an EPS/PDF photo (graphicx package needed) extra braces are
% needed around the contents of the optional argument to biography to prevent
% the LaTeX parser from getting confused when it sees the complicated
% \includegraphics command within an optional argument. (You could create
% your own custom macro containing the \includegraphics command to make things
% simpler here.)
%\begin{IEEEbiography}[{\includegraphics[width=1in,height=1.25in,clip,keepaspectratio]{mshell}}]{Michael Shell}
% or if you just want to reserve a space for a photo:

% \begin{IEEEbiography}{Michael Shell}
% Biography text here.
% \end{IEEEbiography}

% if you will not have a photo at all:
% \begin{IEEEbiographynophoto}{John Doe}
% Biography text here.
% \end{IEEEbiographynophoto}

% insert where needed to balance the two columns on the last page with
% biographies
%\newpage

% \begin{IEEEbiographynophoto}{Jane Doe}
% Biography text here.
% \end{IEEEbiographynophoto}

% You can push biographies down or up by placing
% a \vfill before or after them. The appropriate
% use of \vfill depends on what kind of text is
% on the last page and whether or not the columns
% are being equalized.

%\vfill

% Can be used to pull up biographies so that the bottom of the last one
% is flush with the other column.
%\enlargethispage{-5in}

% that's all folks
\end{document}